\documentclass[preprint,12pt,authoryear]{elsarticle}

\usepackage{siunitx}
\usepackage{hyperref}
\usepackage{float}
\usepackage{subcaption}
\usepackage{xhfill}
\usepackage{xcolor}
\usepackage{tikz}
\usetikzlibrary{arrows}
\tikzstyle{arrow} = [thick,->,>=stealth]
\usepackage{calc}
\usepackage{pgfplots}
\usepackage{pgf}
\usetikzlibrary{positioning}

\usepackage[utf8]{inputenc}
\usepackage{standalone}
\usepackage{breakcites}
\usepackage{microtype}

\usepackage{colortbl}
\usepackage{multirow}

\pgfplotsset{compat=1.16}
 
\usepgfplotslibrary{polar}

\usepackage{pgfplotstable}

\usepackage{pgfplots}

\floatstyle{plaintop}
\restylefloat{table}

\usepackage{tikz}
\usetikzlibrary{arrows}

\usepackage{graphicx}
\usepackage{amssymb}

\begin{document}

\biboptions{semicolon} 

\journal{Journal of Wind Engineering and Industrial Aerodynamics }

\begin{frontmatter}


\title{Influence of Geometry Acquisition Method on Pedestrian Wind Simulations}



\author{Trond-Ola Hågbo, Knut Erik Teigen Giljarhus and Bjørn Helge Hjertager}

\address{University of Stavanger, Department of Mechanical and Structural Engineering and Materials Science, Kjell Arholmsgate 41, Stavanger, Norway}

\begin{abstract}
The construction of a building inevitably changes the microclimate in its vicinity. Many city authorities request comprehensive wind studies before granting a building permit, which can be obtained by Computational Fluid Dynamics (CFD) simulations. When performing wind simulations, the quality of the geometry model is essential. Still, no available studies examine different geometry inputs' impact on the wind flow through an urban environment. This study investigates the influence of the building geometry acquisition method on the simulated wind field in an urban area, focusing on the application of pedestrian wind comfort. A suburban area in the west coast of Norway was chosen as a case study. Four building model types were produced and used in the simulations for comparison. The simulations using a building model produced from data stored in the national general feature catalog (FKB) in Norway showed minor differences to the simulations using more detailed and accurate models based on remote sensing measurements. Prominent variations were seen using a model based on the extrusion of the building footprint. A greater understanding of the geometry acquisition method's influence may enable more efficient pedestrian wind comfort studies that recognize the uncertainty of different geometric model use in urban wind simulations.

\end{abstract}


\end{frontmatter}


\section{Introduction}
\label{S:1}

Computational fluid dynamics (CFD) has been increasingly applied to solve urban wind engineering problems, such as pedestrian wind comfort (\citeauthor{BLOCKEN201215}, \citeyear{BLOCKEN201215}; \citeauthor{BLOCKEN201650}, \citeyear{BLOCKEN201650}; \citeauthor{JANSSEN2013547}, \citeyear{JANSSEN2013547}; \citeauthor{BLOCKEN2009255}, \citeyear{BLOCKEN2009255}) and wind loads on buildings and bridges (\citeauthor{TAMURA20081974}, \citeyear{TAMURA20081974}; \citeauthor{HUANG2007612}, \citeyear{HUANG2007612}; \citeauthor{MONTAZERI2013137}, \citeyear{MONTAZERI2013137}; \citeauthor{THORDAL2019155}, \citeyear{THORDAL2019155}). Knowledge of the wind flow at a given geographic location complements numerous applications in climate modeling, urban planning, environmental modeling, and other applications (\citeauthor{blocken201450}, \citeyear{blocken201450}). 

When performing wind simulations in an urban environment, the quality of the geometry model is essential. Since geometric modeling can be a time-consuming task, and simulations cannot capture all details, the geometry is typically simplified. This simplification comes at a price as it compromises the simulation's ability to describe the local flow behavior it represents.  

The required accuracy of the geometric models and the wind simulations is highly dependent on the application. A higher level of simplification can be accepted for applications only when the main flow structure is of interest and when there is high uncertainty associated with other simulation parameters. An example of such an application may be the assessment of pedestrian wind comfort in an urban environment. 

\subsection{Geometric building models}
In 2017, a review of the CFD analysis of urban microclimates was published. Urban microclimate can be defined as the local climate observed in urban areas, which can be significantly different from those of the surrounding rural areas. The review compared available studies from 1998-2015 (\citeauthor{TOPARLAR20171613}, \citeyear{TOPARLAR20171613}). Among the 183 studies, none of them examined the influence of different geometry inputs on the wind flow through an urban environment. For most of the studies analyzing real urban environments, the buildings were represented in the geometric models by highly simplified blocks, often extrusions of the building footprints, for example \citeauthor{BAIK201248} (\citeyear{BAIK201248}), \citeauthor{TOPARLAR201579} (\citeyear{TOPARLAR201579}), \citeauthor{TAN2016265} (\citeyear{TAN2016265}), and \citeauthor{NG2012256} (\citeyear{NG2012256}). Some studies included highly detailed building data models for a few buildings in question, typically one or two, while the surrounding buildings were simplified blocks, for example \citeauthor{BLOCKEN201215} (\citeyear{BLOCKEN201215}), \citeauthor{BLOCKEN2009255} (\citeyear{BLOCKEN2009255}), \citeauthor{JANSSEN2013547} (\citeyear{JANSSEN2013547}), and \citeauthor{MOONEN2012197} (\citeyear{MOONEN2012197}). 

Detailed building models with highly accurate geometries can be generated through the use of remote sensing techniques, such as photogrammetry and Light Detection and Ranging (LiDAR). Even though these technologies' availability and maturity have greatly improved in recent years, their potential to be used in urban wind simulations remains underutilized. A notable exception is \citeauthor{LUKAC20171} (\citeyear{LUKAC20171}) that used a classified LiDAR point cloud to generate 2.5D models of the buildings. 

While the use of remote sensing enables an enhanced level of detail and accuracy in the building models, it does not necessarily improve the reliability or value of the wind simulations for all applications. It can significantly increase pre-processing time due to the complex meshing and simulation time resulting from the higher number of grid cells necessary to resolve greater detail. Additionally, the process of generating the geometric model itself can be highly time-consuming. Using automatically generated models could make wind simulations significantly more efficient to perform. These are reasons for investigating the influence of geometry acquisition method on pedestrian wind simulations and their associated accuracy. 

New buildings' construction alters the local wind environment in their vicinity, and high-rise buildings are especially influential. They tend to lead high winds down to ground level, which reduces pedestrians' wind comfort and safety.  This effect is a growing concern because of the ongoing urbanization and increasing number of high-rise buildings, as well as climate change resulting in the possibility of more frequent high winds. 
 
Therefore, pedestrian-level wind conditions are often among the first microclimatic issues to be considered in modern city planning and building design (\citeauthor{wu2012designing}, \citeyear{wu2012designing}). Some city authorities request extensive wind analysis before granting a building permit. The possible negative impact to the local wind environment by the erection of the planned buildings can be reduced by making informed design decisions. Wind simulations using CFD can be of great aid in this process.   

\subsection{Aim and main objectives}
This study aims to investigate the influence of geometry acquisition method on the simulated wind field in an urban area. The main application in this study is towards pedestrian wind comfort. The results are also relevant for other applications, such as HVAC (heating, ventilation, and air conditioning) inlets and building pressure evaluation. An improved understanding may help to develop more efficient and reliable tools to perform wind simulations in urban environments.  

Wind simulations using four different types of 3D geometric models of the buildings have been conducted and compared in this study. These are:
\begin{enumerate}
    	\item Footprint extrusion model
        \item FKB database model
        \item Drone photogrammetry model 
        \item Airborne LiDAR model
\end{enumerate}

\begin{figure}[tbp]
    \centering
    \begin{tikzpicture}
    \draw (0, 0) node[inner sep=0] {\includegraphics[width=0.6\textwidth]{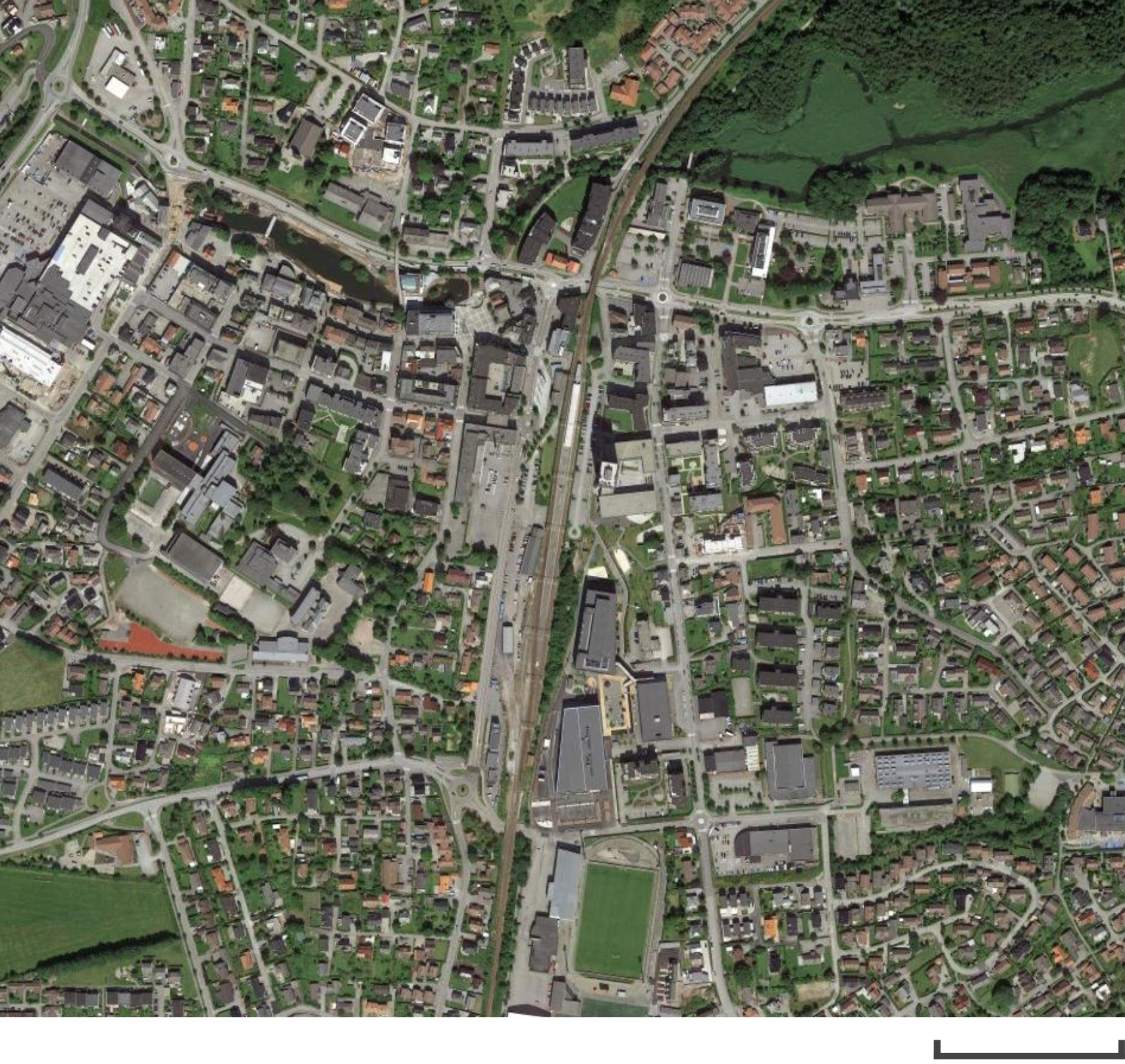}};
    \draw (2,-3.75) node {100 m};
    \end{tikzpicture}
    \caption{Bryne city center \cite{gmaps}.}
    \label{fig:Bryne}
    
\end{figure}

The location chosen for this study is Bryne city center. 
Bryne is located on the Southwest coast of Norway, a region considered to be windy, especially in the fall and winter. It is a suburban/urban area with relatively flat terrain and one distinctive high-rise building, see Figure \ref{fig:Bryne}.

This paper is structured as follows. The next chapter describes the methodology in detail, starting with presenting the geometric building models used in the simulations and briefly the production process. Then, the computational domain and grids used are described together with the numerical setup. The last section of the methodology chapter outlines the process of producing aggregated pedestrian wind comfort maps based on multiple wind simulations, statistical weather data, and defined wind comfort criteria. Finally, the results are presented and discussed.

\section{Methodology}
\label{S:2}

As the accuracy and reliability of CFD simulations can be easily compromised, best practice guidelines (BPGs) have been developed for properly conducting CFD analysis of the wind environment in urban areas. These guidelines are based on the cross-comparison between CFD simulations, wind tunnel experiments, and field measurements on several cases. The BPGs provided by \citeauthor{franke2007cost} (\citeyear{franke2007cost}) and \citeauthor{tominaga2008aij} (\citeyear{tominaga2008aij}) were closely followed to set up the simulations.

\subsection{The geometry models}

The geometric models used for the wind simulations consist of two layers: a \emph{terrain} layer and a \emph{buildings} layer. As this study focuses on the influence of the building models on the urban wind simulations, four building model types are applied and compared, while the same terrain model is used for all simulations. In this section, the procedures of producing the geometric models are briefly explained, and the variations in the building models presented.  

\subsection{The terrain model} 

The terrain model used in the wind simulations is produced from a DTM (digital terrain model) created with multi-band satellite photogrammetry. Well-known remote sensing techniques have been utilized to filter out vegetation, buildings, cars, etc., from the model, leaving only the bare ground. The DTM used to produce the terrain model was provided by The Norwegian Mapping Authority (\citeauthor{Kartverket2},  \citeyear{Kartverket2}).

\subsection{The building models} 

Four building model types are used in the simulations: the Footprint extrusion model, the FKB database model, the Drone photogrammetry model, and the Airborne LiDAR model. They are separated by the level of detail and resemblance to the location they represent. The models' deviations are mostly a result of different sources of technology used to acquire the data. Choices in the model production phase cause minimal changes.

\begin{figure}[tbp]
    \centering
    \begin{subfigure}{0.46\textwidth}
        \includegraphics[width=\textwidth]{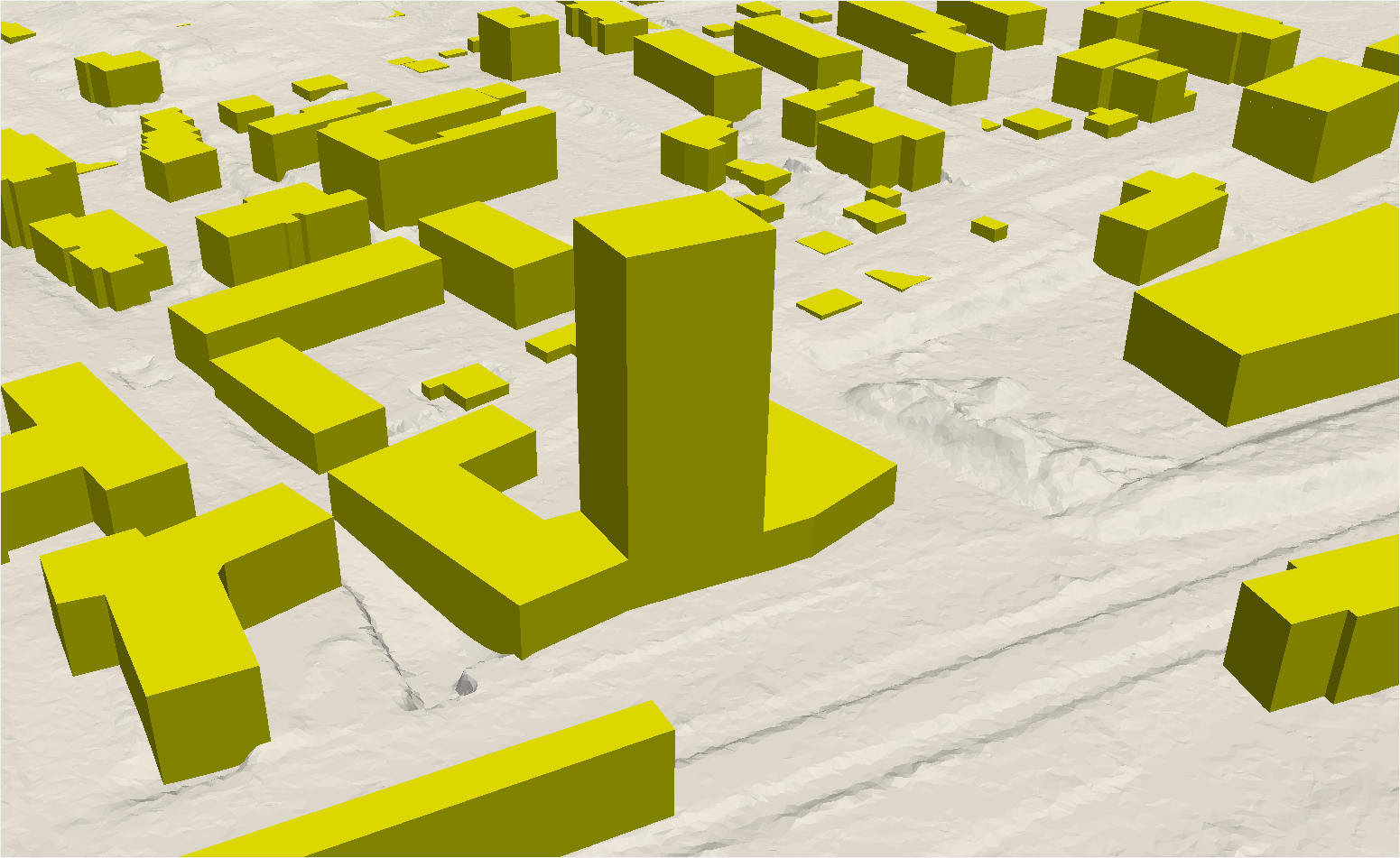}
        \caption{Footprint extrusion model.}
    \end{subfigure}
    \hspace{1pc}
    \begin{subfigure}{0.46\textwidth}
        \includegraphics[width=\textwidth]{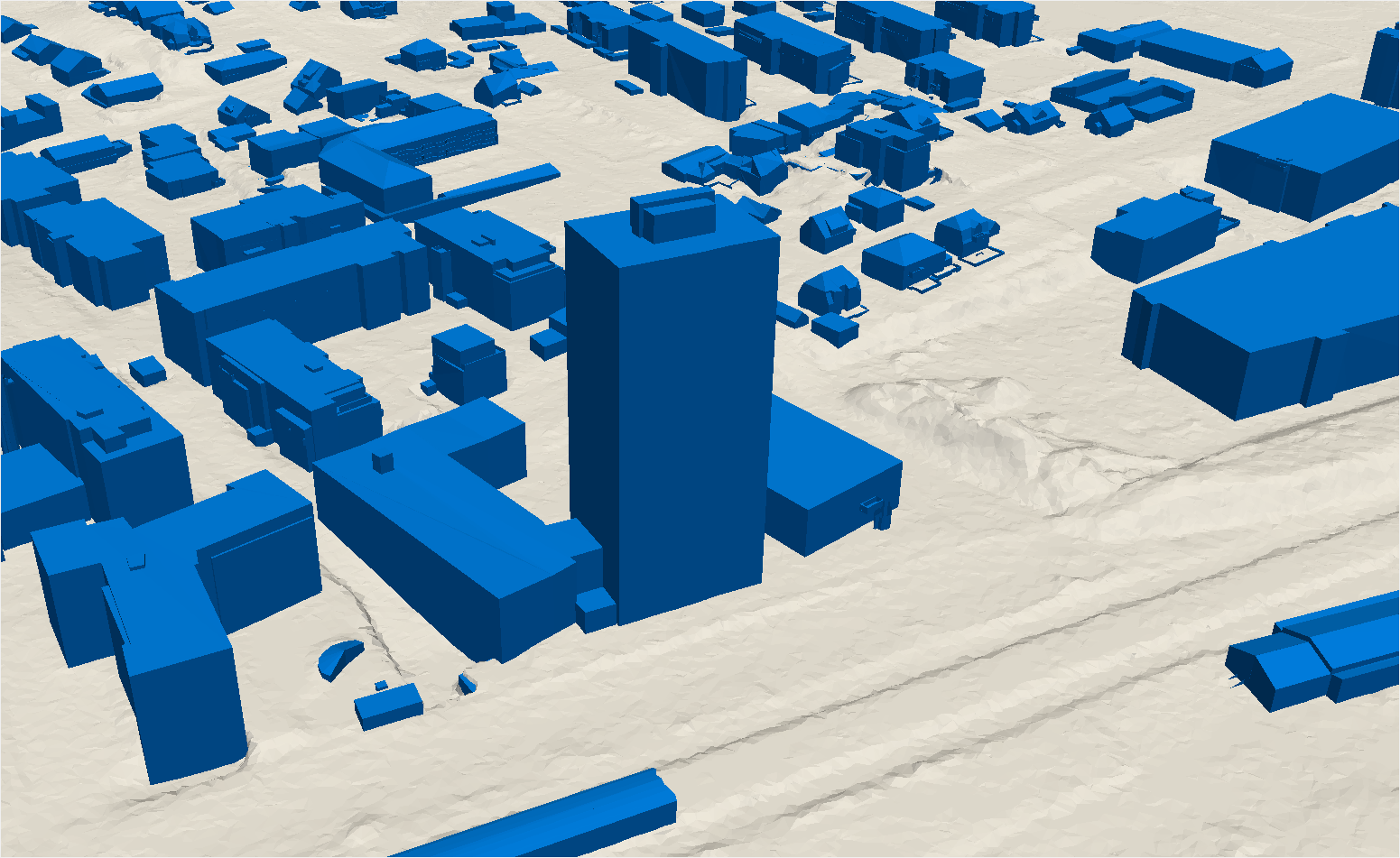}
        \caption{FKB database model.}
    \end{subfigure}

    \vspace{0.7pc}

    \begin{subfigure}{0.46\textwidth}
        \includegraphics[width=\textwidth]{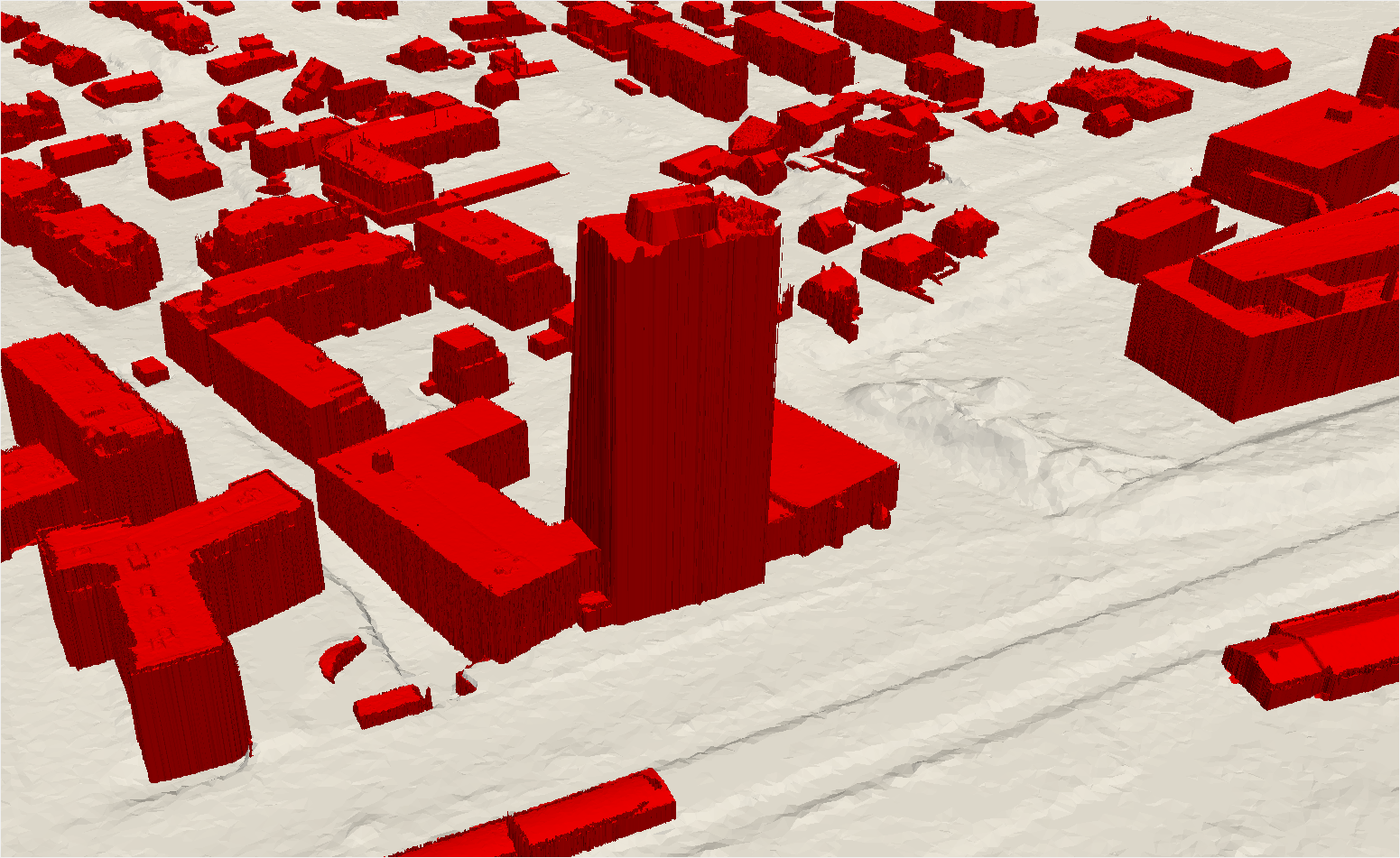}
        \caption{Drone photogrammetry model.}
    \end{subfigure}
    \hspace{1pc}
    \begin{subfigure}{0.46\textwidth}
        \includegraphics[width=\textwidth]{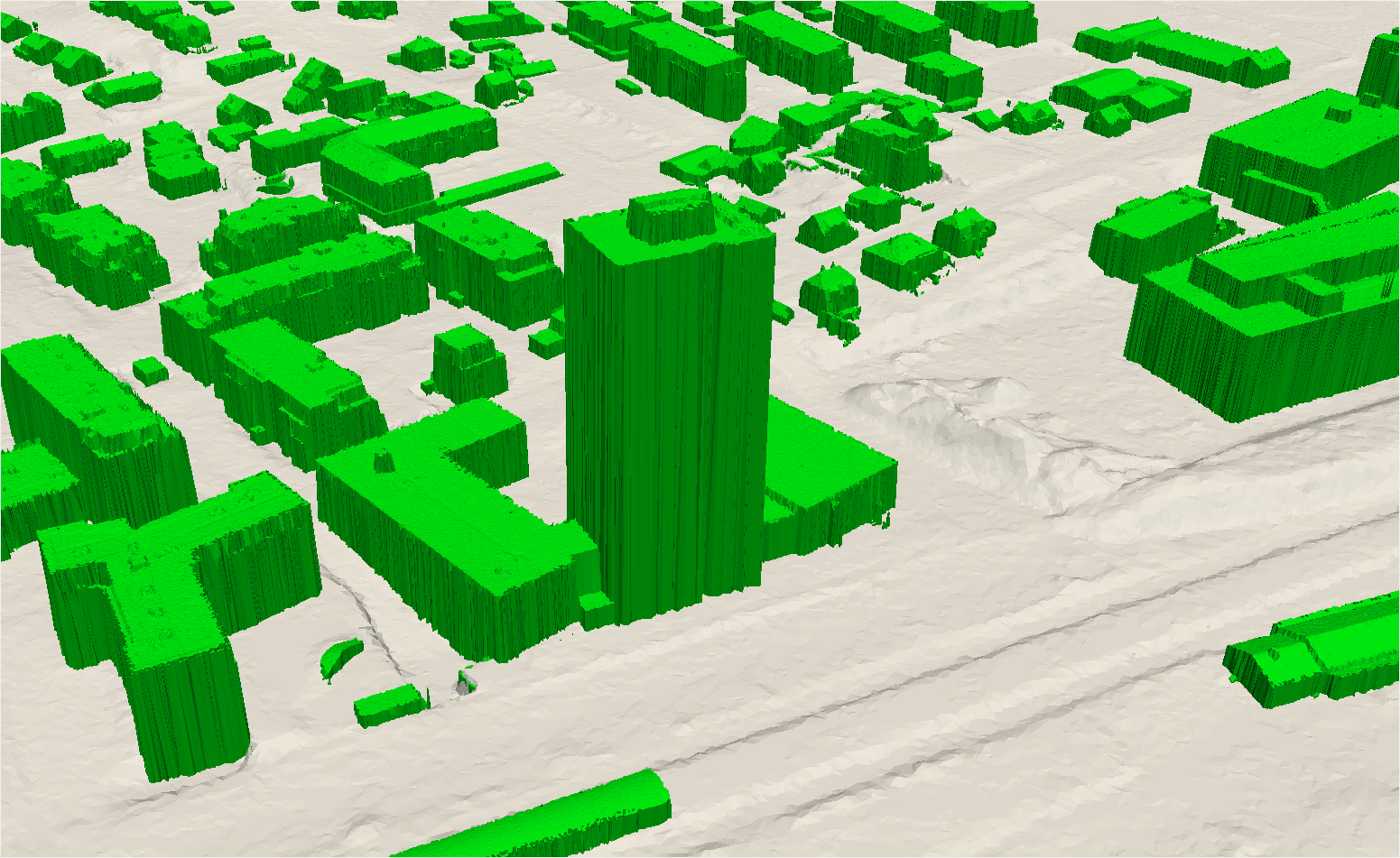}
        \caption{Airborne LiDAR model.}
    \end{subfigure}
    \caption{A comparison of the different building model types, here the high-rise building and surrounding area.}
    \label{fig:buildingmodel}
\end{figure}

Figure \ref{fig:buildingmodel} illustrates some of the variations in the finished building models in a few selected buildings. The next sections describe how the data is acquired, the process of producing wind-simulation-ready geometric models, and present the variations in the models quantitatively.

\subsubsection{Footprint extrusion model} 
This building model type consists of highly simplified building blocks and is commonly seen in urban wind simulations and is the most simplified representation of the buildings used in this study. Examples include \citeauthor{TOPARLAR201579} (\citeyear{TOPARLAR201579}), \citeauthor{TAN2016265} (\citeyear{TAN2016265}), \citeauthor{NG2012256} (\citeyear{NG2012256}), and \citeauthor{tominaga2008aij} (\citeyear{tominaga2008aij}). As the name implies, the building models are extruded from the registered footprint of the buildings. The building height is not set based on measurements, but set according to a city building database or calculated height based on the number of registered floors and standard floor height. 
 
\subsubsection{FKB database model} 
The FKB database model consists of simplified models maintaining general building features. FKB (Felles KartdataBase) is a general feature database governed by the Norwegian national mapping authority. The information used in the model production process is stored in the SOSI (Systematic Organization of Spatial Information) format, a national standard format for geographical information. The required level of detail and precision for the models is highest in urban areas (FKB-A and FKB-B standard) (\citeauthor{Kartverket}, \citeyear{Kartverket}). Most of the building data is based on airborne photogrammetry.

\subsubsection{Drone photogrammetry model} 
Both the Drone photogrammetry model and the Airborne LiDAR model are produced from point cloud data acquired by remote sensing measurements. The finished models are relatively similar, even though the technology used to produce them differ.  

The data used to produce this model is acquired through drone photogrammetry. It works by taking multiple overlapping photos of the ground along a flight path and compares them to determine the elevation height of points in the terrain, buildings, and other objects. The drone sampling of this study's spatial data was only equipped with a single camera taking photos in the visual electromagnetic spectrum, which cannot penetrate vegetation and buildings.

\subsubsection{Airborne LiDAR model} 
The data used to produce this model is acquired through airborne LiDAR(Light Detection and Ranging). LiDAR works by emitting light in the form of a pulsed laser and measuring the time it takes to be reflected and return to a sensor, here mounted on an aircraft.  The distance to the target object is calculated based on the return time as the speed of light is known. LiDAR can penetrate thin/sparse objects, providing multiple returns as various surfaces are 'hit.' The difference in signal intensity is used to classify the points, which can be utilized to exclude undesirable objects.

\subsubsection{Data origin} 
Information on building footprint and height was obtained from OpenStreetMap (\citeauthor{OpenStreetMap}, \citeyear{OpenStreetMap}) -A collaborative project providing access to geodata and maps to the public for free.  Geodata produced the FKB database model used in this study. Geograf AS produced the drone photogrammetry point cloud, and Statens Kartverk provided the LiDAR point cloud.

\subsubsection{Production of the building models} 
For fast and efficient building model generation of the Footprint extrusion model, the information was loaded into Blender (\citeauthor{blender-osm}, \citeyear{blender-osm}) via a plugin called “Blender-osm” (\citeauthor{blender-osm}, \citeyear{blender-osm}). Blender is an open-source 3D computer graphics software toolset. For both the Footprint extrusion model and the FKB database model, minor manual changes were made post model generation using Blender to ensure the models were watertight and wind-simulation ready. It includes adding a few missing building walls, closing narrow gaps between buildings, and removing unwanted objects. 

\begin{figure}[tbp]
    \centering
    \begin{tikzpicture}
    \draw (0, 0) node[inner sep=0] {\includegraphics[width=\textwidth]{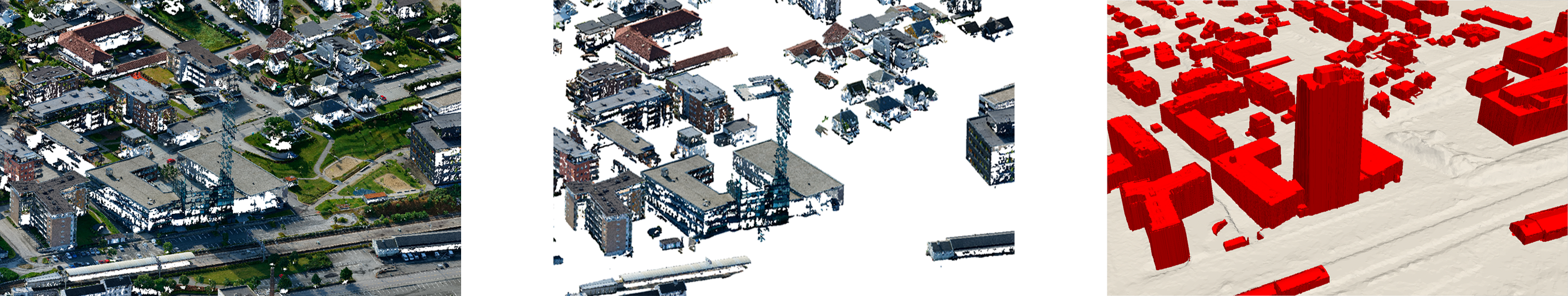}};
    \draw [arrow] (-2.85,0) -- (-2,0);
    \draw [arrow] (2,0) -- (2.85,0);
    \end{tikzpicture}
    \caption{Photogrammetry point cloud to building model process.}
    \label{fig:photogrametry_process}
\end{figure}

\begin{figure}[tbp]
    \centering
    \begin{tikzpicture}
    \draw (0, 0) node[inner sep=0] {\includegraphics[width=\textwidth]{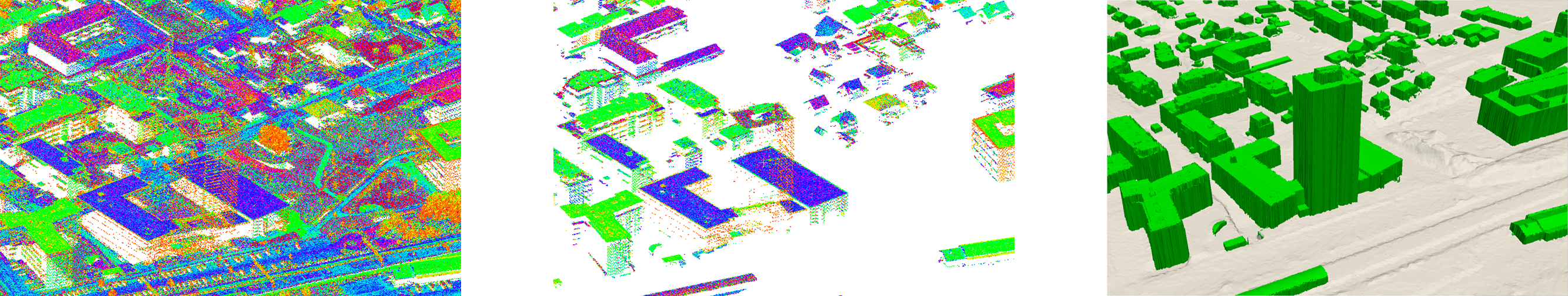}};
    \draw [arrow] (-2.85,0) -- (-2,0);
    \draw [arrow] (2,0) -- (2.85,0);
    \end{tikzpicture}
    \caption{LiDAR point cloud to building model process.}
    \label{fig:lidar_process}
\end{figure}

The process of producing the building models from the point clouds is illustrated in Figure~\ref{fig:photogrametry_process} and Figure \ref{fig:lidar_process}. It involved removing as many undesired points from the point cloud as possible before generating the model. Undesired points represent any other objects than the buildings, such as cars, trees, vegetation, and the ground itself.

\begin{sloppypar}
First, the points representing the ground were filtered out on CloudCompare (\citeauthor{CloudCompare}, \citeyear{CloudCompare}) from the point clouds by applying the Cloth Simulation Filter (CSF) developed by Zhang et al. (\citeauthor{zhang2016easy}, \citeyear{zhang2016easy}). CloudCompare is an open-source point cloud and mesh processing software. Secondly, the remaining undesired points were manually removed by visual inspection. This manual process was time-consuming, especially for the drone photogrammetry point cloud. Some of these undesired points were filtered out from the LiDAR point cloud by exploiting the information of the return signal intensity and number of returns, which significantly reduced the required manual work.
\end{sloppypar}

The building walls were then extruded a few meters downwards to ensure no buildings to be above the terrain layer before generating a surface with 2.5 D Delaunay triangulation. Lastly, Meshlab (\citeauthor{LocalChapterEvents:ItalChap:ItalianChapConf2008:129-136}, \citeyear{LocalChapterEvents:ItalChap:ItalianChapConf2008:129-136}) was utilized for minor cleanup and compression of the model. It is an open-source toolset for processing and editing 3D triangular meshes.

Ongoing development at the location led to the absence of a few buildings in some of the data sets as the spatial data was sampled at different times.  These buildings were removed from all the models to reduce the impact of varying geometry acquisition times.

\subsubsection{Variations in the building models} 
\begin{figure}[tbp]
    \centering
    \begin{subfigure}{0.45\textwidth}
        \includegraphics[width=0.95\textwidth]{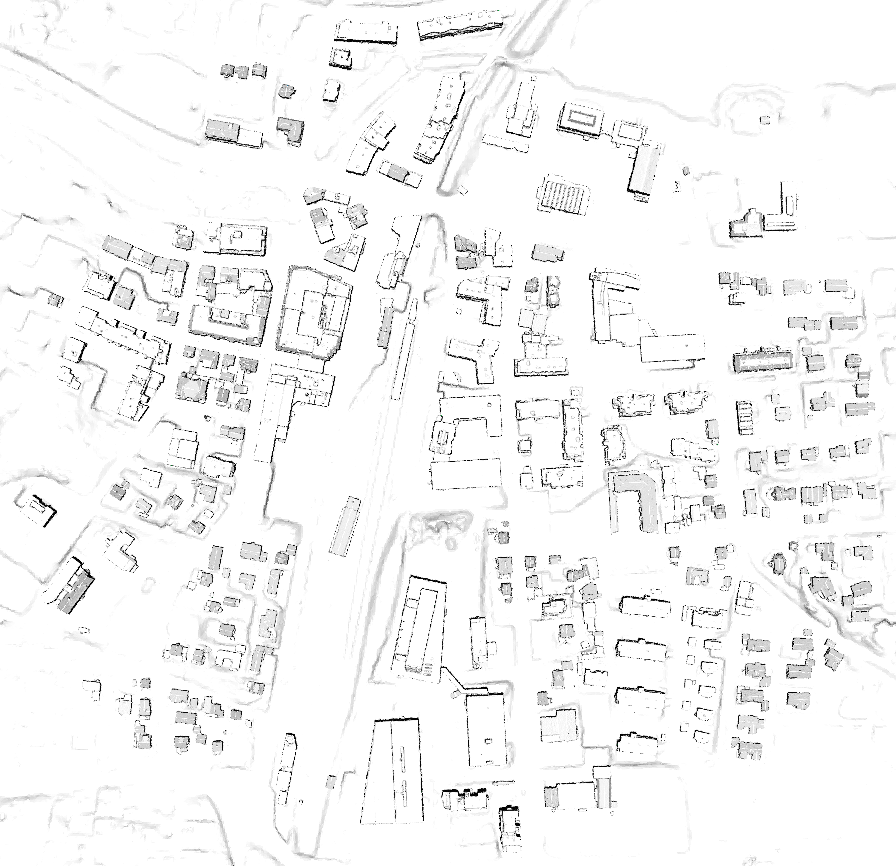}
        \caption{Airborne LiDAR model, reference.}
    \end{subfigure}
    \begin{subfigure}{0.45\textwidth}
        \includegraphics[width=0.95\textwidth]{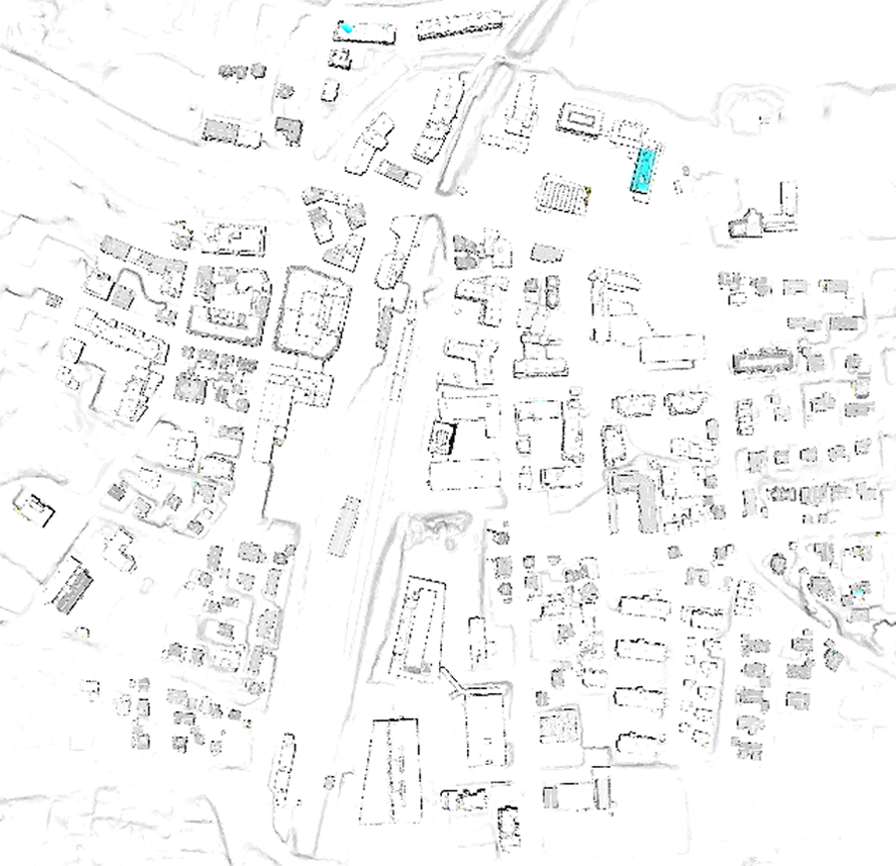}
        \caption{Drone photogrammetry model.}
    \end{subfigure}
    
    \vspace{1pc}
    
    \begin{subfigure}{0.45\textwidth}
        \includegraphics[width=0.95\textwidth]{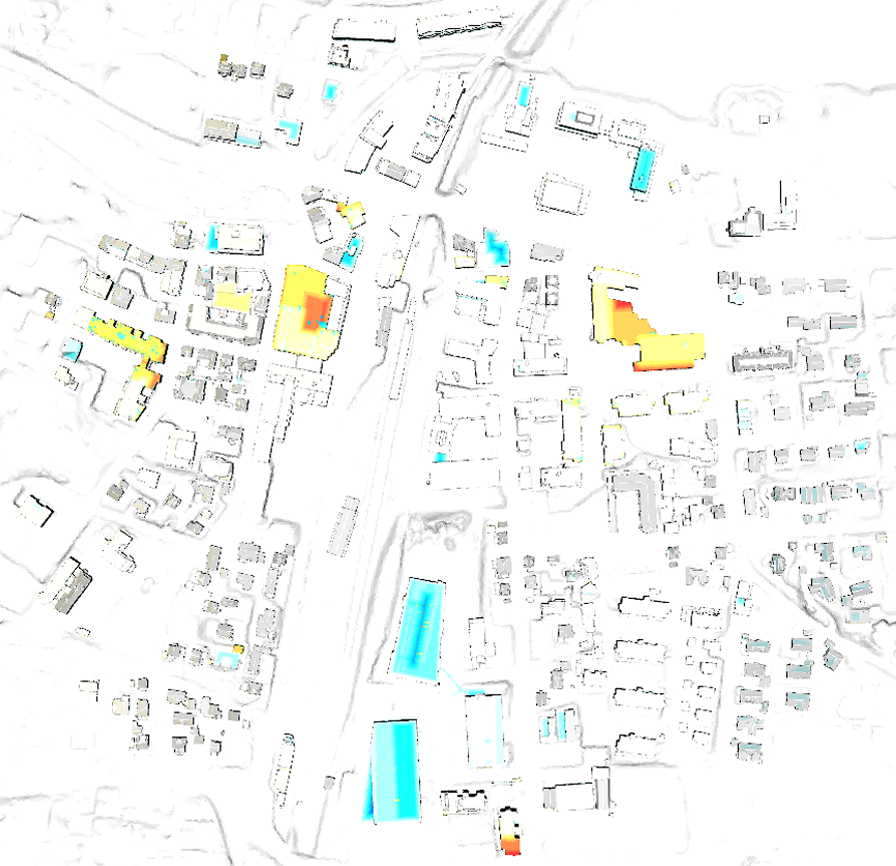}
        \caption{FKB database model.}
    \end{subfigure}
    \begin{subfigure}{0.45\textwidth}
        \includegraphics[width=0.95\textwidth]{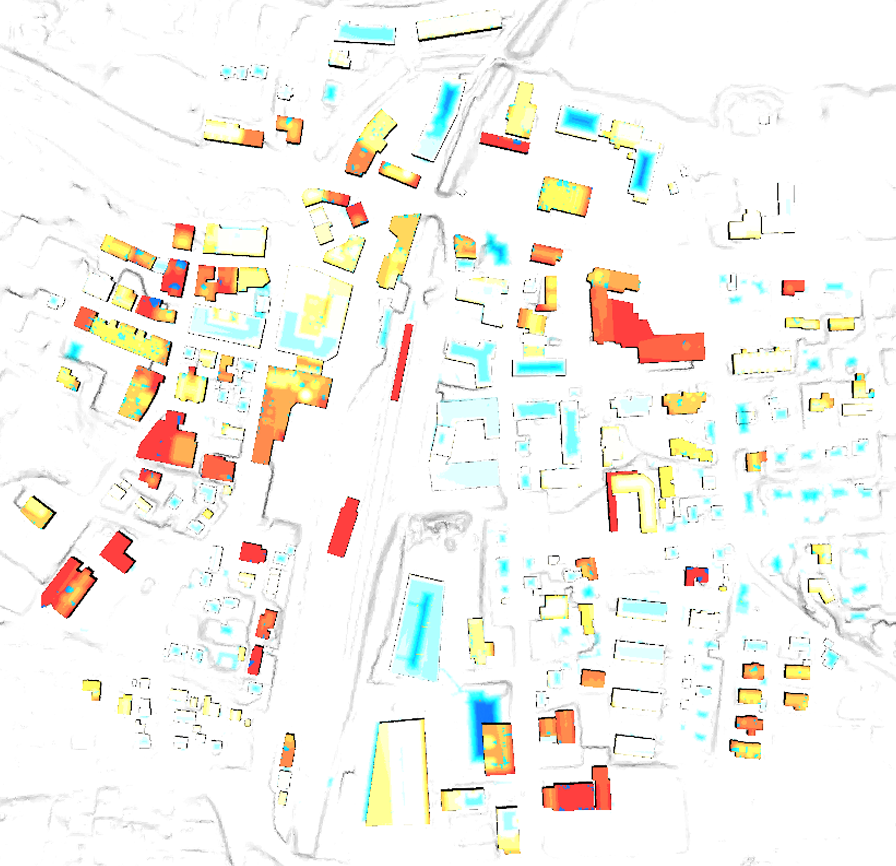}
        \caption{Footprint extrusion model.}
    \end{subfigure}
    
    \vspace{1pc}
    
    \includegraphics[width=0.4\textwidth]{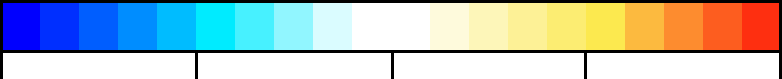}
    
    \begin{tikzpicture}
                \draw (-2.9, 0) node {$-10$};
                \draw (-1.55, 0) node {$-5$};
                \draw (0, 0) node {$0$};
                \draw (1.55, 0) node {\makebox[0pt][l]{$5$}\phantom{$-5$}};
                \draw (2.9, 0) node {\makebox[0pt][l]{$10$}\phantom{$-10$}};
                \draw (0, -0.5) node {$\Delta h$ $(m)$};
        \end{tikzpicture}
    \caption{Top view of the models and the elevation difference, $\Delta h$, to the Airborne LiDAR model.}
    \label{fig:diff}
\end{figure}

A top view of the building models is presented in Figure \ref{fig:diff}. The colors quantify how the models differ in elevation from the Airborne LiDAR model. The Airborne LiDAR model was chosen as a reference as it is likely the most accurate model due to the superior filtering options of LiDAR. Yellow and red represent regions that are more elevated in the models compared to the reference model. Likewise, indigo and blue represent areas in the building models that are lower than the reference.

The Drone photogrammetry model resembles the Airborne LiDAR model the most, and the Footprint extrusion model is the furthest off. Keep in mind that the threshold for visualizing the variations is set to \SI[separate-uncertainty = true]{1(1)}{\meter}. There are minor differences within this threshold in the models that do not appear in the figure.

There is a single building with a high discrepancy in the Drone photogrammetry model compared to the Airborne LiDAR model. It is a  result of different data sampling dates as a garage had been upgraded in the meantime. Most of the buildings on the FKB database model match the buildings' main structure and height in the measurement-based models quite well, despite holding far less information than the point clouds. The Footprint extrusion model matches the buildings' base quite well, but not the height due to lack of accurate information in the OpenStreetMap database.

\subsection{Computational domain}
The BPGs have been used as a minimum requirement when sizing the computational domain. The guidelines are based on box-shaped domains only suitable for wind at the inlet to have one specific direction, which is inward parallel to the domain's sidewalls. As the intention of this study was to have flexible meshes suitable for simulating wind from any horizontal directions, cylindrical domains were used. Therefore, the recommendations of the BPGs had to be applied and adapted to them. 

The routine of using a cylindrical computational domain when performing urban wind simulations was introduced by \citeauthor{kastner2020cylindrical} (\citeyear{kastner2020cylindrical}). They reported that a cylindrical domain yielded comparable results to a traditional box-shaped domain in terms of accuracy and convergence behavior.

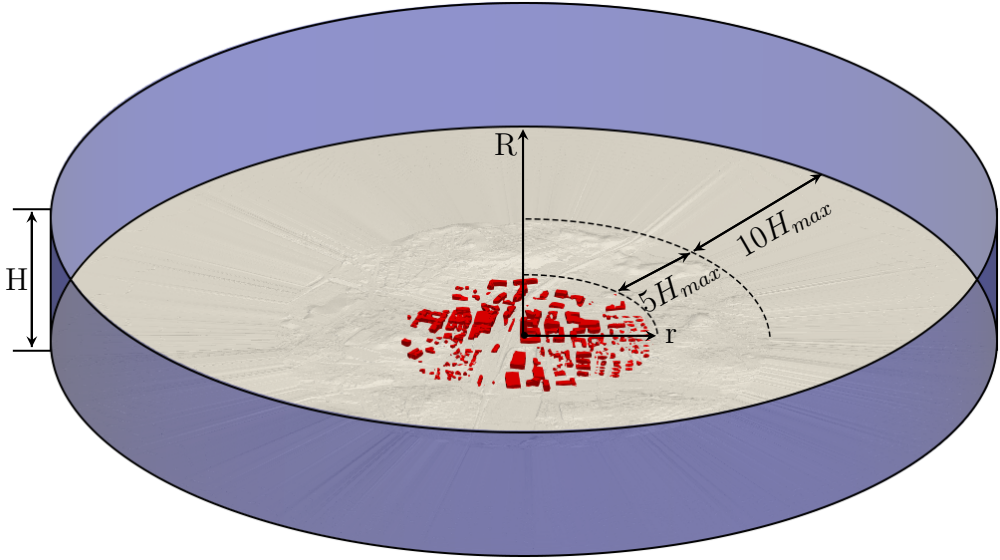
\begin{figure}[tbp]
    \centering
    \begin{tikzpicture}
    \draw (0, 0) node[inner sep=0] {\includegraphics[width=\textwidth]{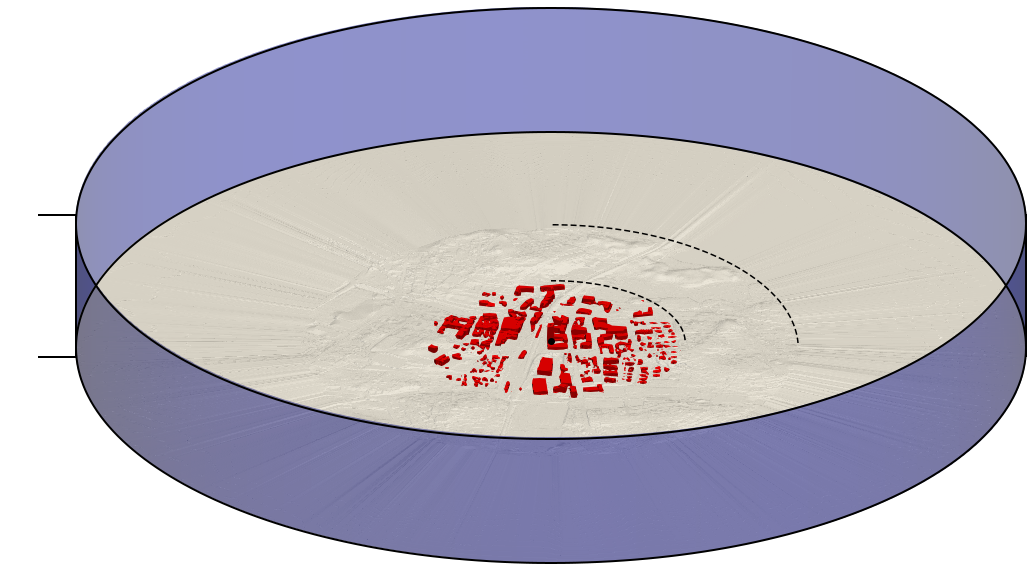}};
    \draw [arrow] (0.42,-0.75) -- (2.2,-0.75);
    \draw [arrow] (0.43,-0.75) -- (0.43,2);
    \draw [arrow] (-6.1,-0.9) -- (-6.1,0.9);
    \draw [arrow] (-6.1,0.9) -- (-6.1,-0.9);
    \draw [arrow] (1.7,-0.15) -- (2.65,0.35);
    \draw [arrow] (2.65,0.35) -- (1.7,-0.15);
    \draw [arrow] (2.7,0.38) -- (4.4,1.4);
    \draw [arrow] (4.4,1.4) -- (2.7,0.38);
    \draw (-6.3,0) node {H};
    \draw (2.4,-0.75) node {r};
    \draw (0.2,1.8) node {R};
    \draw (3.25,0.35) node [rotate=30] {$5 H_{max}$ \enskip $10 H_{max}$};
    \end{tikzpicture}
    \caption{Computational domain.}
    \label{fig:domain}
    
\end{figure}

Figure \ref{fig:domain} presents the computational domain and the dimensions are listed in Table \ref{tab:domain}.

    \begin{table}[tbp]
    \centering
    \begin{tabular}{l l l}
    \hline
    \textbf{Parameter} & \textbf{Length}\\
    \hline
    $H_{max}$  & \SI{66}{\meter}\\
    $\Delta h_t$  & \SI{37}{\meter} \\
    r  & \SI{400}{\meter} \\
    R & \SI{400}{\meter} + $(15 \times H_{max}) =$ \SI{1390}{\meter}\\
    H & \SI{15}{\meter} + $(6 \times H_{max}) =$  \SI{411}{\meter} \\
    \hline
    \end{tabular}
    \caption{Dimensions of computational domains in meters.}
    \label{tab:domain}
    \end{table}

The BPGs suggest that the top of the domain should be 5 $H_{max}$ from the top of the tallest building. $H_{max}$, as defined by \citeauthor{franke2006recommendations} (\citeyear{franke2006recommendations}), is the height of the tallest building and is central in sizing the computational domain.  Additionally, \citeauthor{sorensen2012fine} (\citeyear{sorensen2012fine}) recommend using a domain height of at least ten times the difference in the height of the terrain, $\Delta h_t$, when simulating wind flow over complex terrains. Both requirements are satisfied for the computational domain.

\citeauthor{franke2007cost} (\citeyear{franke2007cost}) suggest a distance of 15 $H_{max}$ downstream of the buildings to the outlet of the domain. As the cylindrical domain is suitable for wind coming in from any direction, the total radial distance from the buildings to the domain's sides is set to 15 $H_{max}$. The first 5 $H_{max}$ includes the terrain, while the outermost 10 $H_{max}$ is a smoothed transition from terrain to flat.  The transition enables the application of realistic boundary conditions at the sides while attaining converged solutions in the wind simulations. Also, including an extended section upstream of the buildings allows the flow to develop and adapt to the terrain, thus attaining a more realistic profile before reaching the buildings. 

\subsection{Computational grids}
The computational grids were developed according to the BPGs. A total of four computational grids, one for every building model type, are used in the simulations presented in this study. 

The same setup for the grid generation is used for all the grids. All the final grids are produced with \texttt{snappyHexMesh}, the hex-dominant unstructured mesh generator of OpenFOAM.

\texttt{snappyHexMesh} requires a base mesh and the triangulated surface geometries it conforms the mesh to fit, here the building models and the terrain models. \citeauthor{Hagbo2019} (\citeyear{Hagbo2019}) concluded that an unstructured mesh generated with \texttt{snappyHexMesh} could be sufficient in urban wind simulations where only the general flow features are of interest.

The base meshes are generated with \texttt{blockMesh}, a meshing utility in OpenFOAM suitable for simple structured meshes with grading and curved edges. In the base mesh, the grid is stretched vertically with an overall expansion ratio of three. It means that the width of the top cells is three times the bottom cells. 

In the final meshes, the grids are refined successively near the terrain and building, with eight grid cells between each refinement level. Three refinement levels are used on the terrain patch, while one additional level is applied for the building patch.  

All the grids consist of about 35 million cells. The resolution is sufficiently high according to the BPGs as it satisfies the recommendation of a minimum of ten cells per building side and in the passage between buildings, as well as at least three cells from the ground to the elevation height (\citeauthor{franke2007cost}, \citeyear{franke2007cost}; \citeauthor{tominaga2008aij}, \citeyear{tominaga2008aij}; \citeauthor{yoshie2007cooperative}, \citeyear{yoshie2007cooperative}). In this study, the elevation height was set to \SI{2}{\meter} above ground level, a reasonable height for pedestrian wind comfort evaluation. Additionally, a mesh sensitivity study was performed to ensure that the grid resolution is sufficiently high. 

\begin{figure}[tbp]
    \centering
        \includegraphics[width=0.8\textwidth]{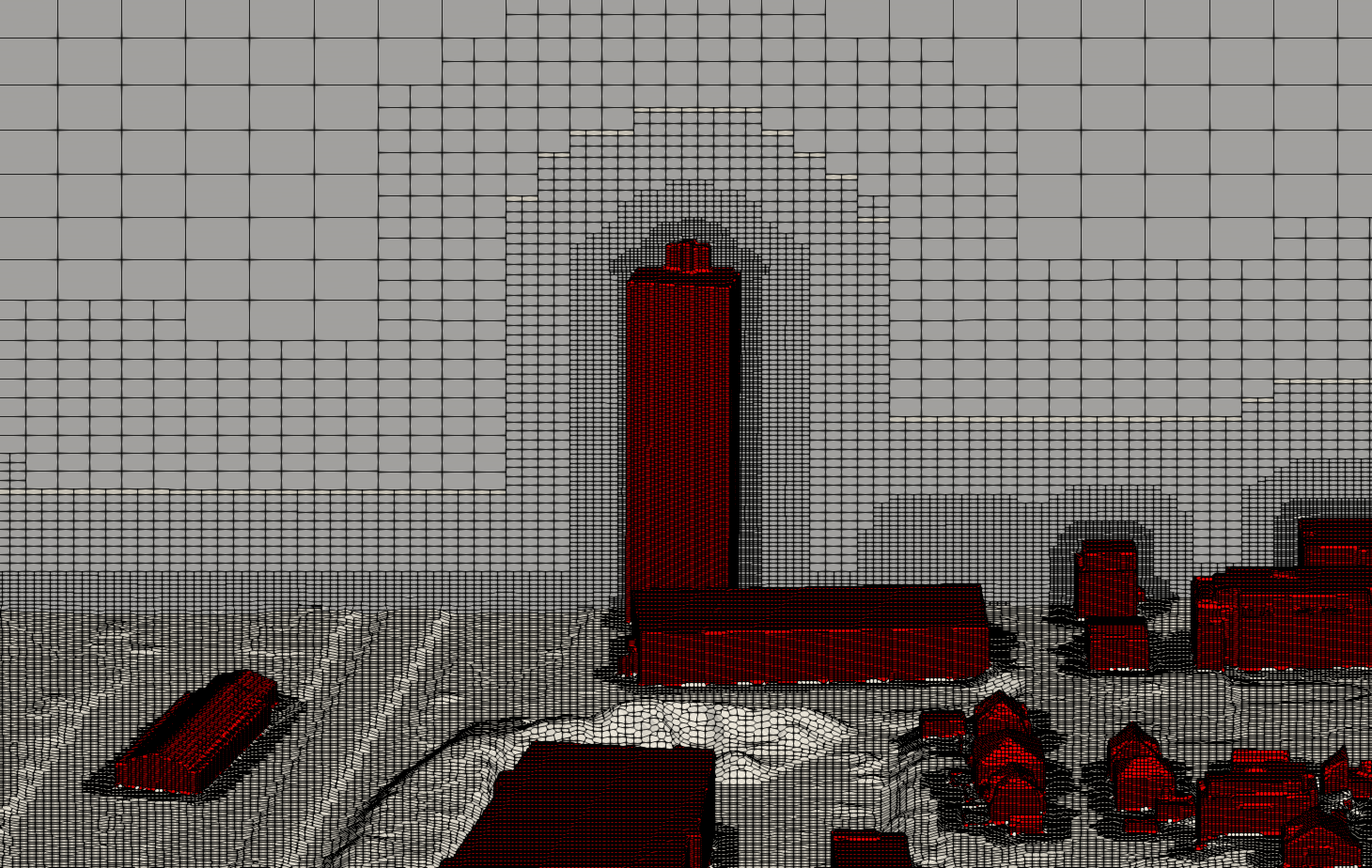}
    \caption{Computational grid structure, partial view from the side.}
    \label{fig:grid}
\end{figure}

Figure \ref{fig:grid} illustrates the grid structure used in all the wind simulations inside the domain from the side, exemplified in the FKB database model mesh.

\subsection{Simulation setup and other computational parameters}
 
The wind simulations were performed by solving the incompressible, three-dimensional steady Reynolds-averaged Navier-Stokes (RANS) equations with the finite volume method. Turbulent closure was provided by the realizable k-$\epsilon$ turbulence model developed by \citeauthor{shih1995new} (\citeyear{shih1995new}) and the SIMPLE algorithm handled the pressure-velocity coupling ensuring the continuity equation to be satisfied. Second-order discretization schemes were used for the spatial discretization, including the convective terms. The simulations were conducted using the OpenFOAM toolbox, which is an open-source CFD software package (\citeauthor{weller1998tensorial}, \citeyear{weller1998tensorial}). 

An overview of the boundary conditions (BCs) used is presented in Table \ref{tab:BC}. Combined inlet/outlet boundary conditions (Robin conditions) are used on the computational domain's sides to allow for wind in any direction. The inlet profiles were set to form a homogeneous atmospheric boundary layer (ABL) with neutral stratified conditions, according to the BPGs. The wind speed at a reference height of 10 m, $U_{ref10m}$, was set to 5 m/s. At the outlet, outlet conditions were used with constant pressure and zero gradient for the remaining variables.  

No-slip conditions with wall functions were used for both the terrain and the building layer. The logarithmic law for smooth walls was used for the buildings. A rough wall condition was applied for the terrain (\citeauthor{hargreaves2007use}, \citeyear{hargreaves2007use}). The roughness was set to $z_0=0.05$, representing grassland. The slip condition was applied to the top boundaries.

\begin{table}
\centering
\begin{tabular}{l l}
\hline
\textbf{Patch} & \textbf{Boundary condition type}\\
\hline
Sides & Inlet/outlet   \\
Inlet  & Atmospheric boundary layer  \\
Outlet  & Fixed pressure and zero gradient  \\
Terrain  & Rough wall, $z_0 = 0.05$  \\
Buildings  & Smooth wall   \\
Top  & Slip  \\

\hline
\end{tabular}
\caption{Overview of boundary conditions applied}
\label{tab:BC}
\end{table}

\subsection{Production of pedestrian wind comfort map}

A pedestrian wind comfort map illustrates the annual wind conditions experienced at ground level by pedestrians at a site. A way of producing these maps is to combine wind simulations, statistical weather data, and a set of defined comfort criteria. There are numerous sets of wind comfort criteria, and in this study, a version of the Lawson wind comfort criteria (\citeauthor{lawson1978widn}, \citeyear{lawson1978widn}) is used, see Table \ref{tab:ped}. The Lawson criteria classify areas with "comfortable" activity types based on their associated accepted average wind speed intervals and exceedance frequency.

\begin{table}
\centering
\begin{tabular}{l l l l l}
\hline
\textbf{Mean wind} & \textbf{Exceedance} &  \textbf{Activity} & \textbf{Class} & \textbf{Color}\\
\textbf{speed range} & \textbf{frequency} &  \textbf{type} &  \textbf{number}\\
\hline

\SIrange[range-units = brackets,open-bracket = {[}]{0}{4}{\meter\per\second} & $\leq$ 5 \% & Sitting & 1 & \cellcolor{blue}\\
\SIrange[range-units = brackets,open-bracket = {[}]{4}{6}{\meter\per\second} & $\leq$ 5 \% & Standing & 2 & \cellcolor{cyan}\\
\SIrange[range-units = brackets,open-bracket = {[}]{6}{8}{\meter\per\second} & $\leq$ 5 \% & Strolling & 3 & \cellcolor{yellow}\\
\SIrange[range-units = brackets,open-bracket = {[}]{8}{10}{\meter\per\second} & $\leq$ 5 \% & Business walking & 4 & \cellcolor{orange}\\
\SI{> 10}{\meter\per\second} & N/A & Uncomfortable & 5 & \cellcolor{red}     \\

\hline
\end{tabular}
\caption{Pedestrian wind comfort criteria used in study, based on Lawson wind comfort criteria\cite{lawson1978widn}.}
\label{tab:ped}
\end{table}

The procedure of producing pedestrian wind comfort maps in this study is as follows. First, the wind is simulated from eight different horizontal directions, including both the cardinal and the intercardinal directions. Then each cell of the computational mesh is categorized according to the comfort criteria based on the wind simulations, and statistical weather data in the form of a wind rose.

\begin{figure}[tbp]
    \centering
        \begin{tikzpicture}
            \makeatletter
            \pgfplotsset
            {
            polar bar/.style=
                {
                scatter,
                draw=none,
                mark=none,
                visualization depends on=rawy\as\rawy,
                area legend,
                legend image code/.code=
                    {%
                    \fill[##1] (0cm,-0.1cm) rectangle (0.6cm,0.1cm);
                    },
                    /pgfplots/scatter/@post marker code/.add code={}
                {
                \pgfmathveclen{\pgf@x}{\pgf@y}
                \edef\radius{\pgfmathresult}
                \fill[]
                    (\pgfkeysvalueof{/data point/x},-\pgfkeysvalueof{/data point/y})
                    ++({\pgfkeysvalueof{/data point/x}-#1/2},\pgfkeysvalueof{/data point/y})
                    arc [start angle=\pgfkeysvalueof{/data point/x}-#1/2,
                        delta angle=#1,
                        radius={\radius pt}
                    ]
                    -- +({\pgfkeysvalueof{/data point/x}+#1/2},-\rawy)
                    arc [start angle=\pgfkeysvalueof{/data point/x}+#1/2,
                        delta angle=-#1,
                        radius={
                            (\pgfkeysvalueof{/data point/y} - \rawy + 0.00001)  / (\pgfkeysvalueof{/data point/y}+ 0.01) * (\radius pt + 0.01)
                                }
                        ]
                    --cycle;
                 }
                },
                polar bar/.default=
            }
                    \begin{polaraxis}
                        [xtick={0,45,...,315},
                        xticklabels={E,NE,N,NW,W,SW,S,SE},
                        ytick={10,20},
                        yticklabels={10\%,20\%},
                        legend cell align={left},
                        legend entries={\,0 to \SI{4}{\meter\per\second}, \,4 to \SI{6}{\meter\per\second}, \,6 to \SI{8}{\meter\per\second}, \,8 to \SI{10}{\meter\per\second}},
                        cycle list={blue, cyan, yellow, orange}
                        ]
                        \pgfplotsinvokeforeach{1,...,4}
                        {
                        \addplot +[polar bar=17, stack plots=y]
                        table [x expr=-\thisrowno{0}+90, y index=#1] {Figures/Methodology/frequencies.csv};
                        }
                    \end{polaraxis}
        \end{tikzpicture}
        \caption{Modified wind rose at \SI{10}{\meter} above ground level, Bryne.}
        \label{fig:windrose1}
    \end{figure}
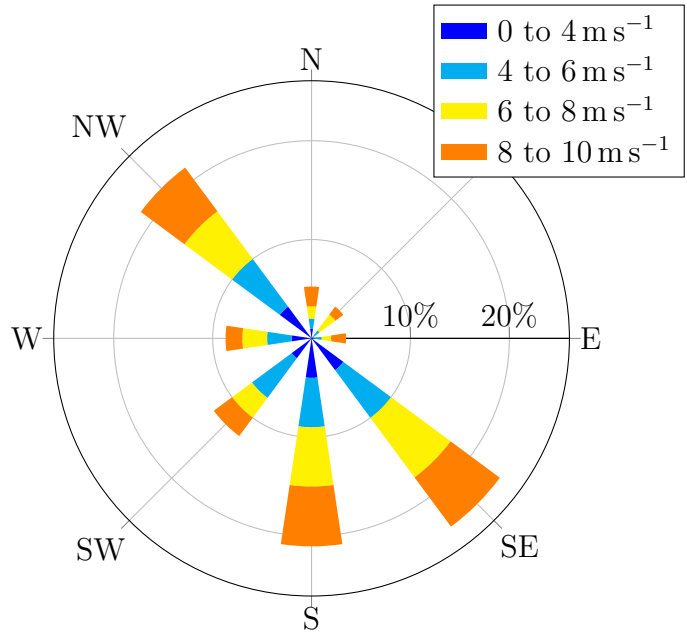

Figure \ref{fig:windrose1} presents the wind rose used in the production of the maps. Note that it divides the observed wind measurements into the same eight wind directions applied at the inlet in the wind simulations.  The wind rose is based on historical wind measurements from the area, but altered to have stronger winds. It was done to enable the usage of all classification classes providing more data for the comparison of the maps.

The wind simulations match the wind roses' wind directions, but not all the wind speeds. Therefore, the wind simulations' resulting scalars are scaled linearly to fit each wind speed interval class's middle before categorizing each cell. The scaling is justified as the wind is well within the turbulent flow regime for all the wind speed intervals under consideration, and hence similar flow distributions are expected. 

Lastly, the classification of the cells located \SI{2}{\meter} above the terrain level is read, and its representing color is projected onto the terrain patch below. The result is a pedestrian wind comfort map.

\subsection{Quantitative image comparison using structural similarity}
The structural similarity index measure (SSIM), proposed by \citeauthor{SSIM} (\citeyear{SSIM}), was introduced to compare the wind simulations' overall similarity at the pedestrian-level. SSIM measures the similarity between two images and works by comparing local patterns of pixel intensities for luminance and contrast. The maximum SSIM value of 1 is only reachable when comparing two identical photos, while 0 indicates no structural similarity. The minimum SSIM value is -1 and corresponds to the comparison of an image with its negative. 

The SSIM was used to measure different building geometry models' influence on the simulated wind velocity at pedestrian-level. It includes wind simulations from eight different directions and the aggregated pedestrian wind comfort maps. All the fields to be compared were projected onto the same reference mesh (the Airborne LiDAR model mesh) before calculating the SSIM. It was done to reduce the impact of different building cutouts in the images.

\section{Results}
\label{S:3}
\subsection{Wind speed at pedestrian level}

\begin{figure}[tbp]
    \centering
    \begin{subfigure}{0.45\textwidth}
        \includegraphics[width=0.95\textwidth]{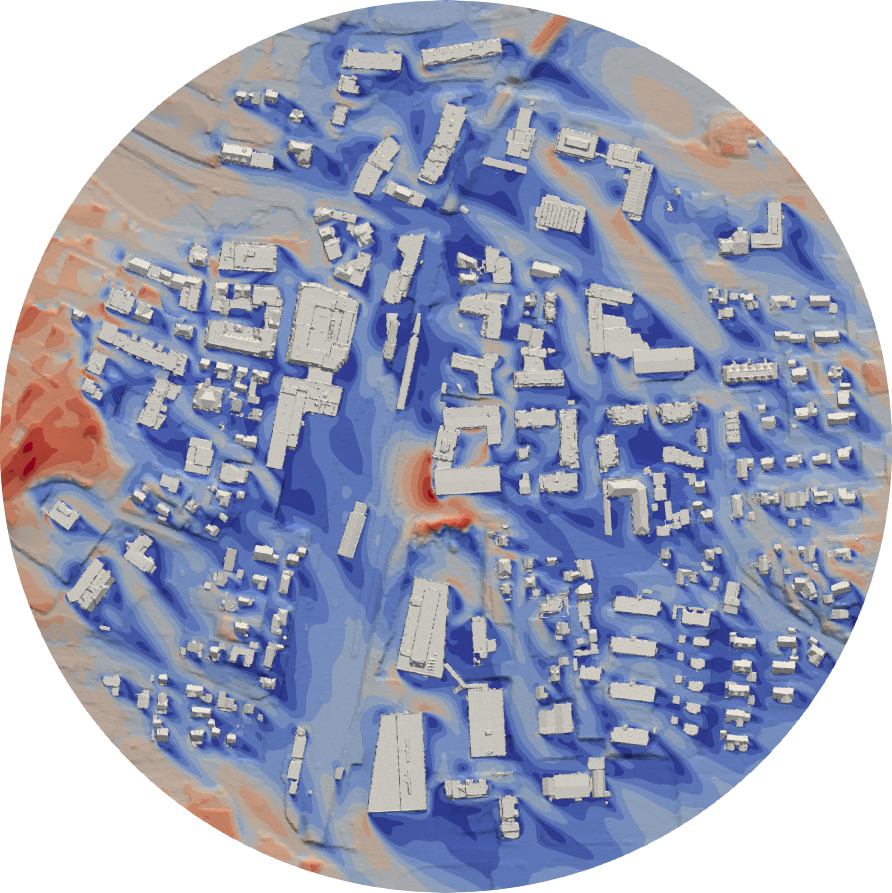}
        \caption{Airborne LiDAR model.}
    \end{subfigure}
    \begin{subfigure}{0.45\textwidth}
        \includegraphics[width=0.95\textwidth]{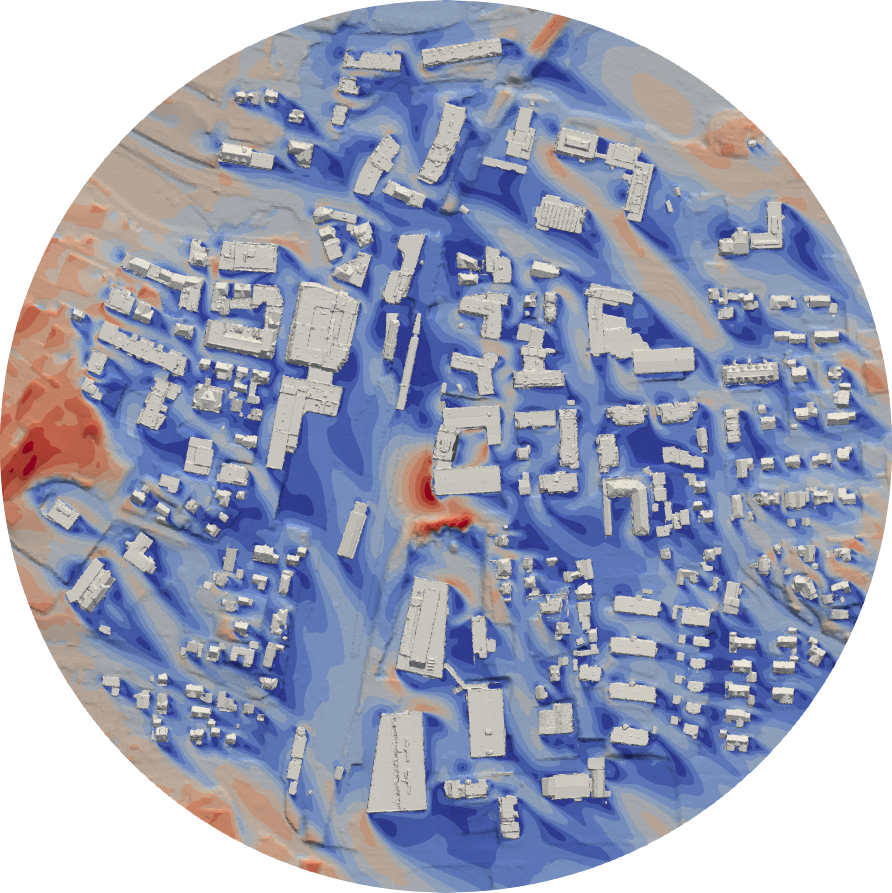}
        \caption{Drone photogrammetry model.}
    \end{subfigure}
    
    \vspace{1pc}
    
    \begin{subfigure}{0.45\textwidth}
        \includegraphics[width=0.95\textwidth]{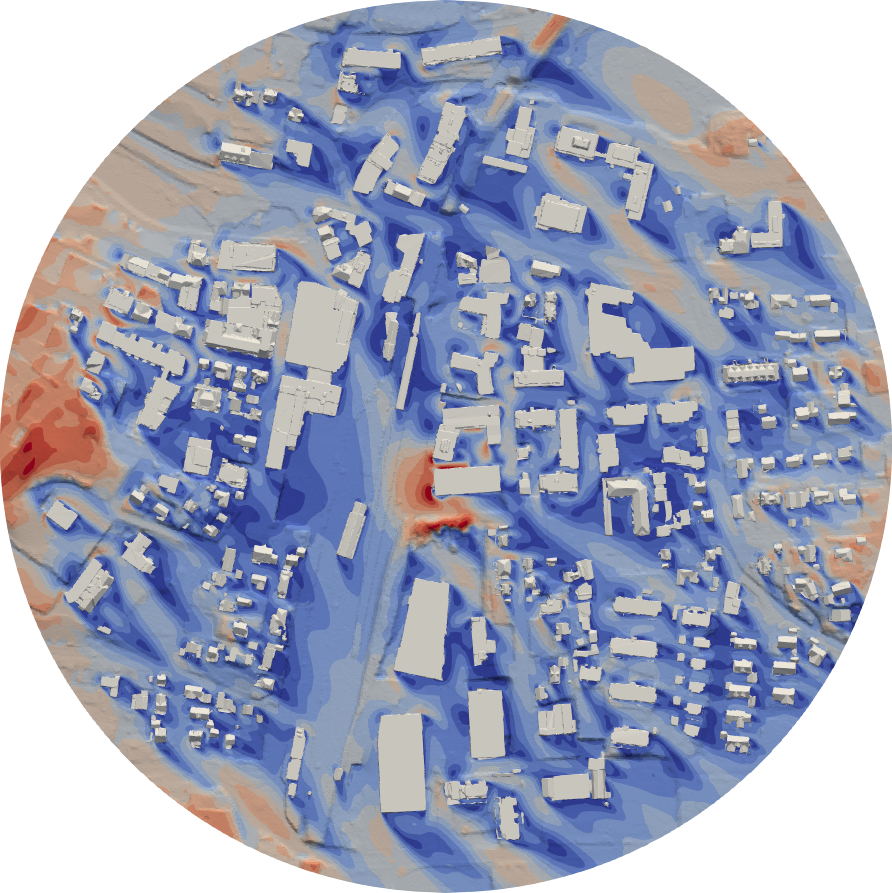}
        \caption{FKB database model.}
    \end{subfigure}
    \begin{subfigure}{0.45\textwidth}
        \includegraphics[width=0.95\textwidth]{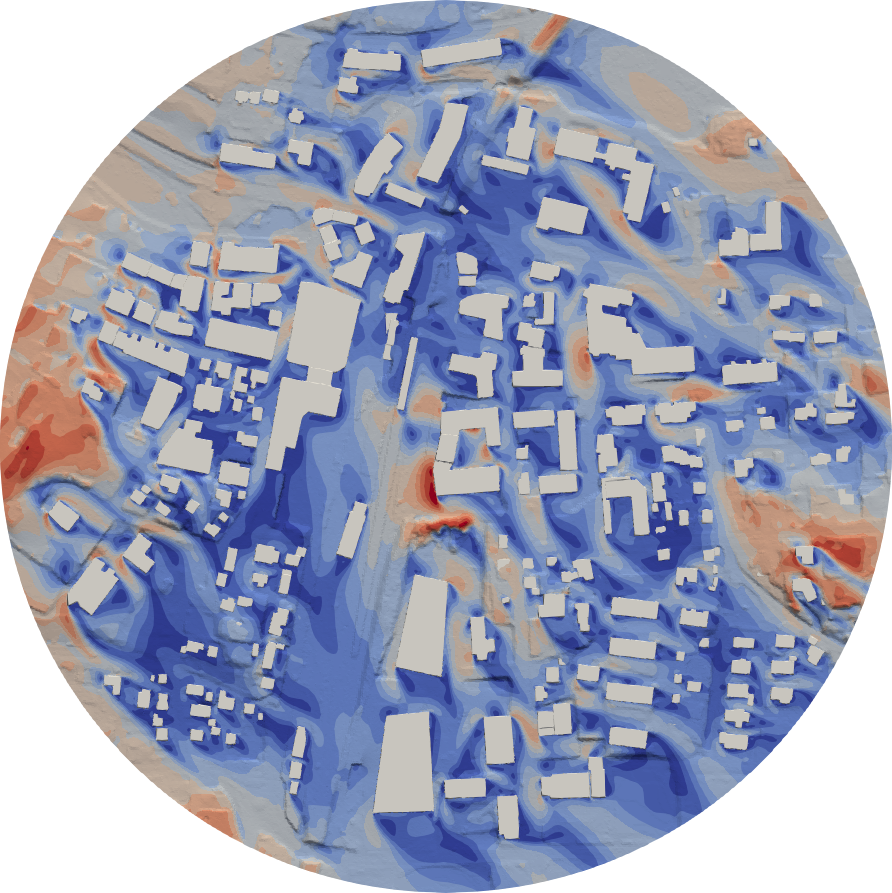}
        \caption{Footprint extrusion model.}
    \end{subfigure}
    
    \vspace{1pc}
    
    \includegraphics[width=0.4\textwidth]{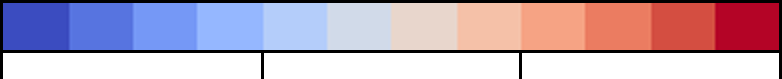}
    
    \begin{tikzpicture}
                \draw (-2.75, 0) node {\makebox[0pt][l]{ $0$}\phantom{$0.6$}};
                \draw (-0.92, 0) node {$0.4$};
                \draw (0.92, 0) node {$0.8$};
                \draw (2.75, 0) node {$1.2$};
                \draw (0, -0.5) node {$U / U_{ref}$};
        \end{tikzpicture}
    \caption{Wind velocity magnitude at \SI{2}{\meter} above ground, wind from NW ($315^{\circ}$).}
    \label{fig:magnitude}
\end{figure}

Figure \ref{fig:magnitude} shows the wind velocity magnitude at \SI{2}{\meter} above ground with the wind coming from Northwest (315$^{\circ}$) relative to the wind velocity at \SI{10}{\meter} elevation at the inlet, $U_{ref}$. Wind from Northwest was specifically chosen to be presented as it illustrates the possible negative impact of a high-rise building on the local wind conditions at pedestrian-level. 

The high-rise building is positioned in the middle of the figures. One of the building side walls faces the wind blowing from the Northwest almost perpendicularly. A result of this is that some of the high-velocity wind from high elevations being directed downwards along the building wall towards the ground, increasing the pedestrian level's wind velocity in the vicinity. 

\begin{figure}[tbp]
    \centering
    \begin{subfigure}{0.45\textwidth}
        \includegraphics[width=0.95\textwidth]{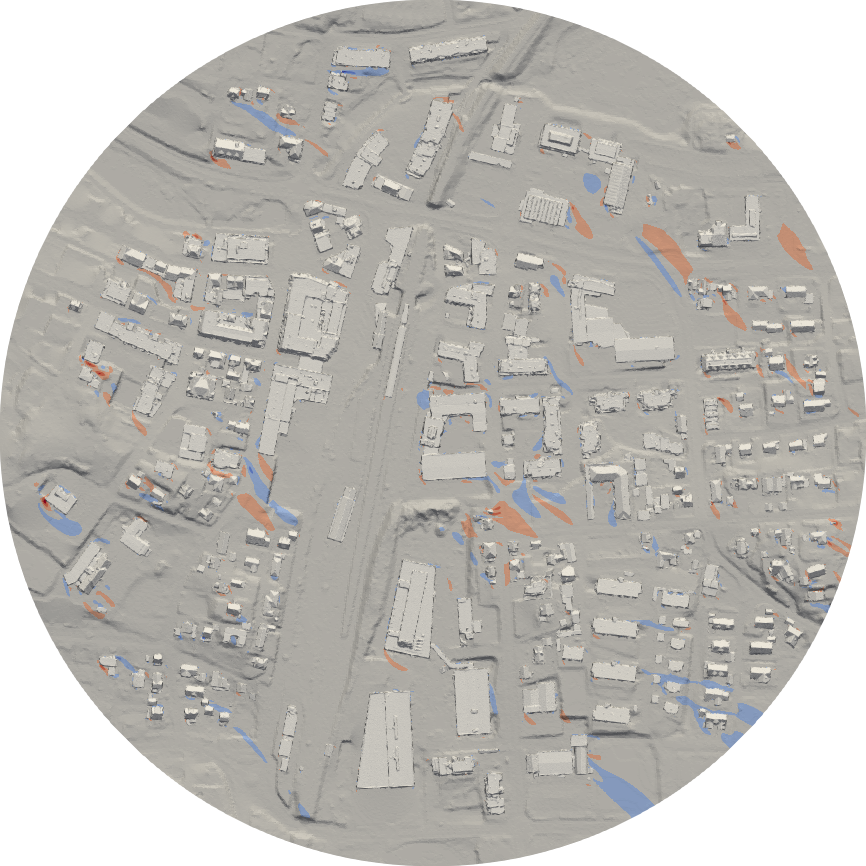}
        \caption{Drone photogrammetry model.}
    \end{subfigure}
    \begin{subfigure}{0.45\textwidth}
        \includegraphics[width=0.95\textwidth]{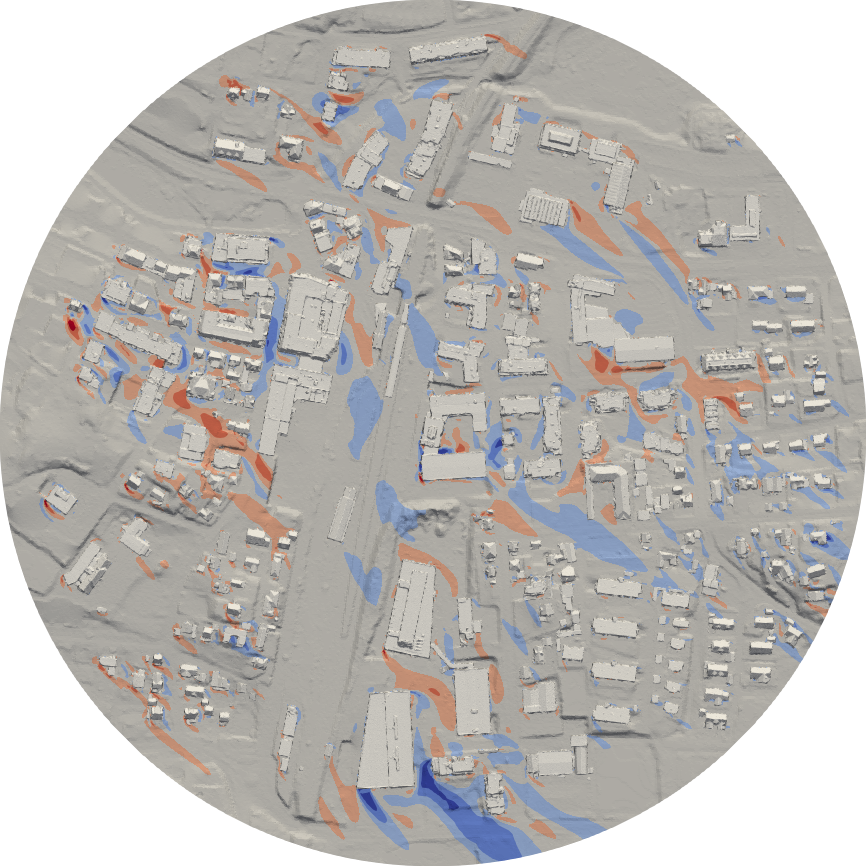}
        \caption{FKB database model.}
    \end{subfigure}
    
    \vspace{1pc}
    
    \begin{subfigure}{0.45\textwidth}
        \includegraphics[width=0.95\textwidth]{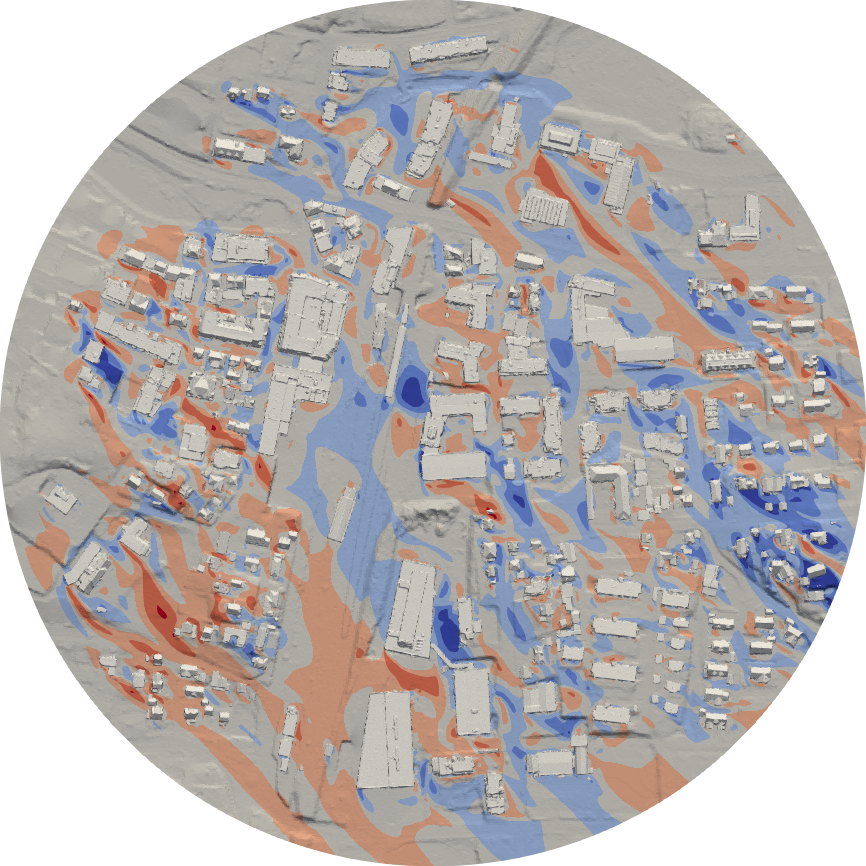}
        \caption{Footprint extrusion model.}
    \end{subfigure}
    
    \vspace{1pc}
    
    \includegraphics[width=0.4\textwidth]{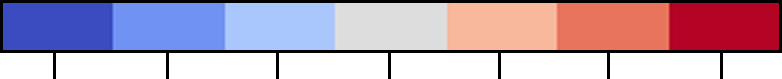}
    
    \begin{tikzpicture}
                \draw (-2.5, 0) node {$-0.6$};
                \draw (0, 0) node {$0$};
                \draw (2.5, 0) node {\makebox[0pt][l]{$0.6$}\phantom{$-0.6$}};
                \draw (0, -0.5) node {$\Delta U / U_{ref}$};
        \end{tikzpicture}
    \caption{Velocity magnitude difference, $\Delta$U, to the reference model at \SI{2}{\meter} above ground, wind from NNW (315$^{\circ}$).}
    \label{fig:magnitudediff}
\end{figure}

The general flow pattern and wind velocity appear to be relatively similar using the different models in the simulations. The differences are best viewed in Figure \ref{fig:magnitudediff}. It shows the difference between the LiDAR model simulation used as a reference and simulations using the other geometry models. Negative values, represented by blue, correspond to areas where the reference case has a lower wind velocity magnitude than the compared case and vice versa.

The Drone photogrammetry model simulation is the most similar to the reference case, and the Footprint extrusion model simulation the least. The extent of the deviations coincides with the similarity to the reference geometry model itself. The variations are more significant near and downstream of the buildings with the most discrepancies to the reference model, see Figure \ref{fig:buildingmodel}. These variations diminish further downstream, but for the Footprint extrusion model simulations they are so severe that the effect is evident further downstream. Similar differences were also seen for the other seven wind directions.

Table \ref{tab:ssim} presents the SSIM intervals of the wind velocity magnitude at \SI{2}{\meter} above ground level from eight wind directions compared to the reference case. Recall that an SSIM value of 1 is only reachable when comparing two identical images, while 0 indicates no structural similarity.  

The table confirms the variations seen in Figure \ref{fig:magnitude} and Figure \ref{fig:magnitudediff} quantitatively. Compared to the reference case, pedestrian-level wind velocity distributions' structural similarities are highest when using the Drone photogrammetry model. Not far behind are the simulations using the FKB database model. The SSIM values indicate that the structural similarity to the reference case is by far the least when comparing to the wind simulations using the Footprint extrusion model. 

Notice the values' consistency when comparing for all the wind directions represented by the relatively narrow intervals. It indicates that the figures' structural similarities are more dependent on the building model used in the wind simulation than the wind direction.

\begin{table}
\centering
\begin{tabular}{l l}
\hline
\textbf{Case compared to reference} & \textbf{SSIM intervall}\\

\hline
Drone photogrammetry model & \SIrange[range-units  = brackets, open-bracket = [, close-bracket= ]]{0.92}{0.94}{}\\
FKB database model & \SIrange[range-units  = brackets, open-bracket = [, close-bracket= ]]{0.89}{0.90}{}\\
Footprint extrusion model & \SIrange[range-units  = brackets, open-bracket = [, close-bracket= ]]{0.82}{0.84}{}\\

\hline
\end{tabular}
\caption{Structural similarity of pedestrian-level wind velocity distributions for eight wind directions to reference simulations.}
\label{tab:ssim}
\end{table}

\subsection{Pedestrian wind comfort aggregated map}
\begin{figure}[tbp]
    \centering
    \begin{subfigure}{0.45\textwidth}
        \includegraphics[width=0.95\textwidth]{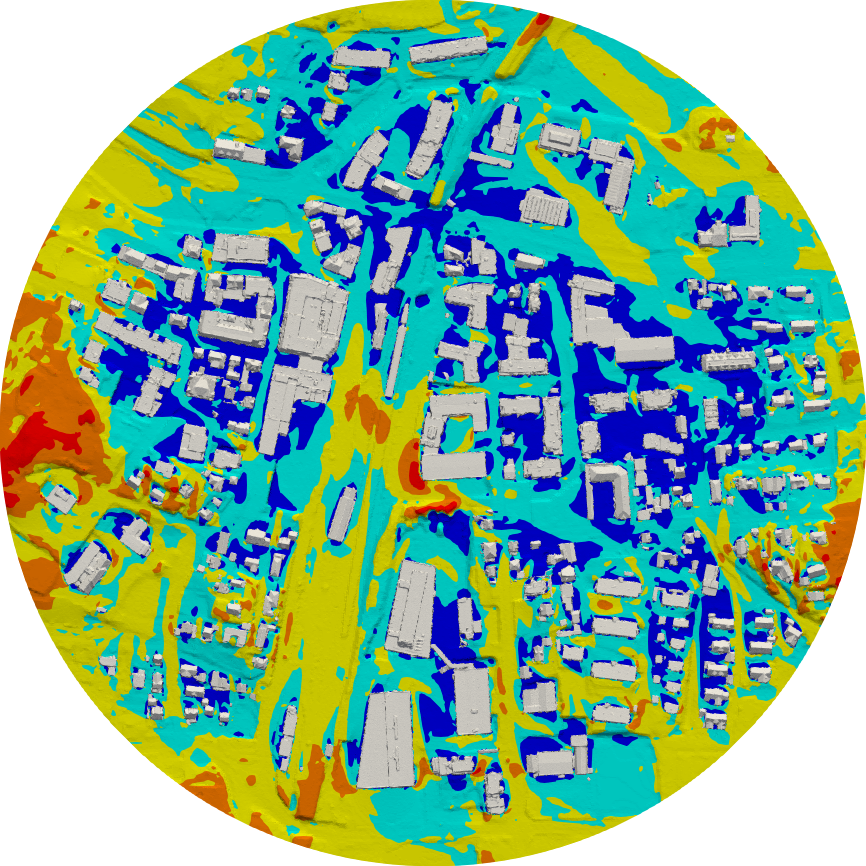}
        \caption{Airborne LiDAR model.}
    \end{subfigure}
    \begin{subfigure}{0.45\textwidth}
        \includegraphics[width=0.95\textwidth]{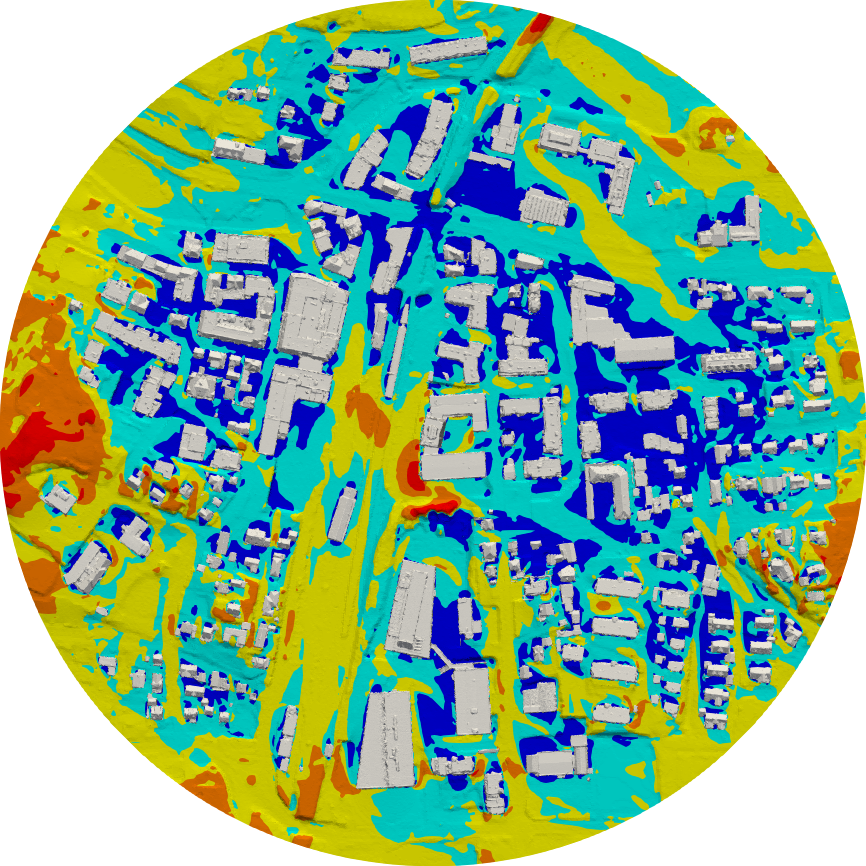}
        \caption{Drone photogrammetry model.}
    \end{subfigure}
    
    \vspace{1pc}
    
    \begin{subfigure}{0.45\textwidth}
        \includegraphics[width=0.95\textwidth]{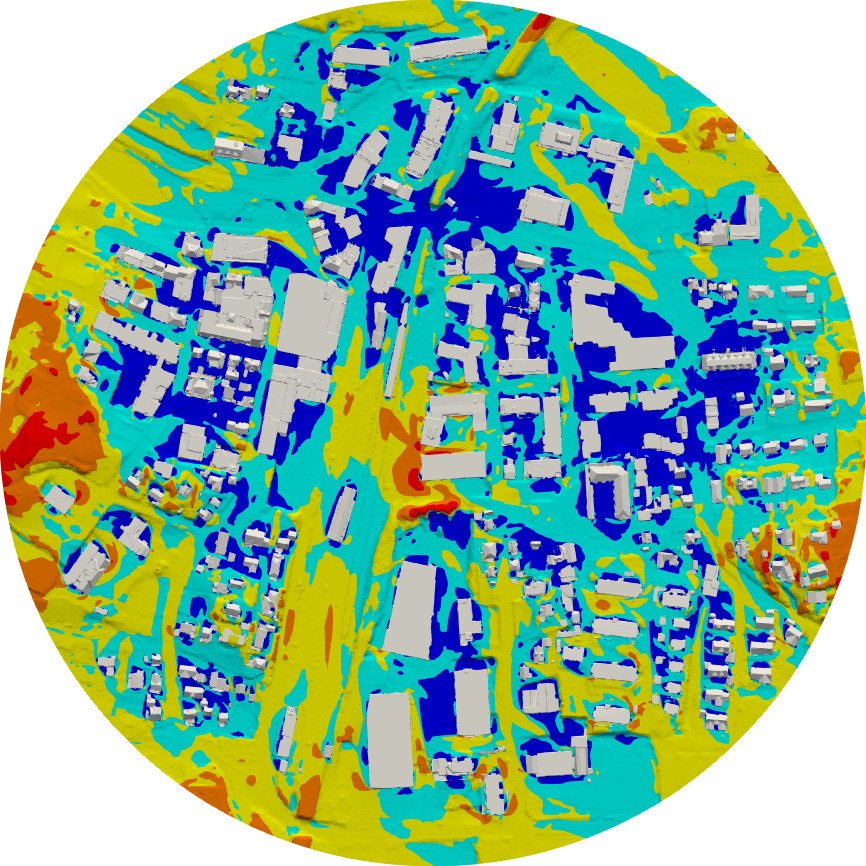}
        \caption{FKB database model.}
    \end{subfigure}
    \begin{subfigure}{0.45\textwidth}
        \includegraphics[width=0.95\textwidth]{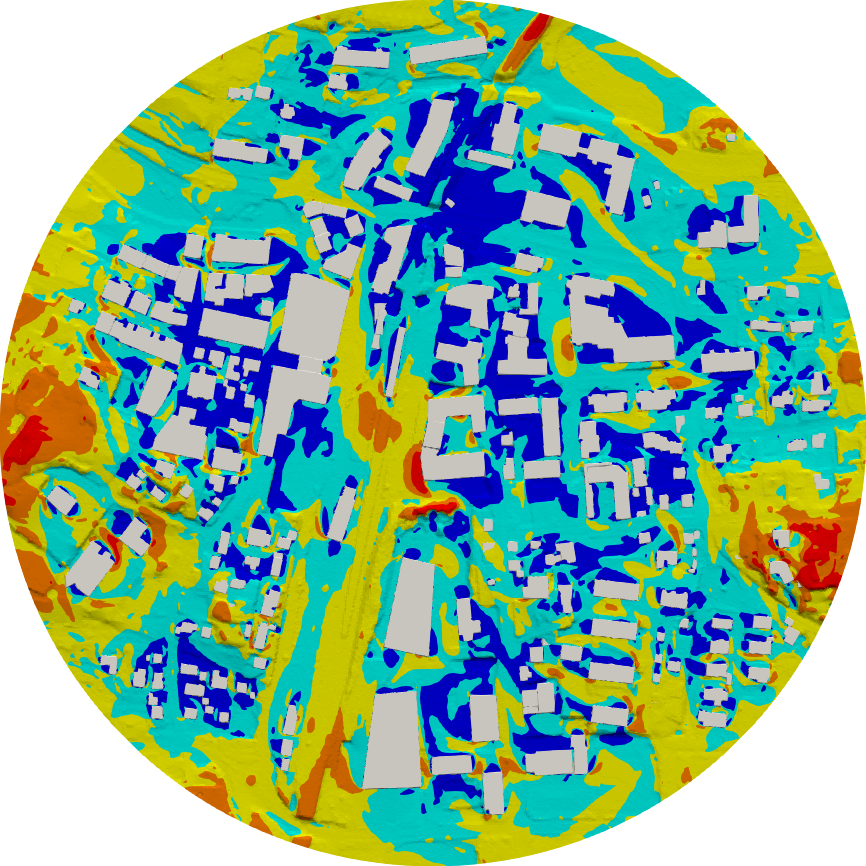}
        \caption{Footprint extrusion model.}
    \end{subfigure}
    
    \vspace{1pc}
    
    \includegraphics[width=0.4\textwidth]{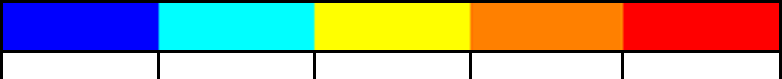}
    
    \begin{tikzpicture}
                \draw (-2.2, 0) node {$1$};
                \draw (-1.1, 0) node {$2$};
                \draw (0, 0) node {$3$};
                \draw (1.1, 0) node {$4$};
                \draw (2.2, 0) node {$5$};
                \draw (0, -0.5) node {Comfort class};
        \end{tikzpicture}
    \caption{Pedestrian wind comfort maps at \SI{2}{\meter} above ground.}
    \label{fig:cproj}
\end{figure}

\begin{figure}[tbp]
    \centering
    \begin{subfigure}{0.45\textwidth}
        \includegraphics[width=0.95\textwidth]{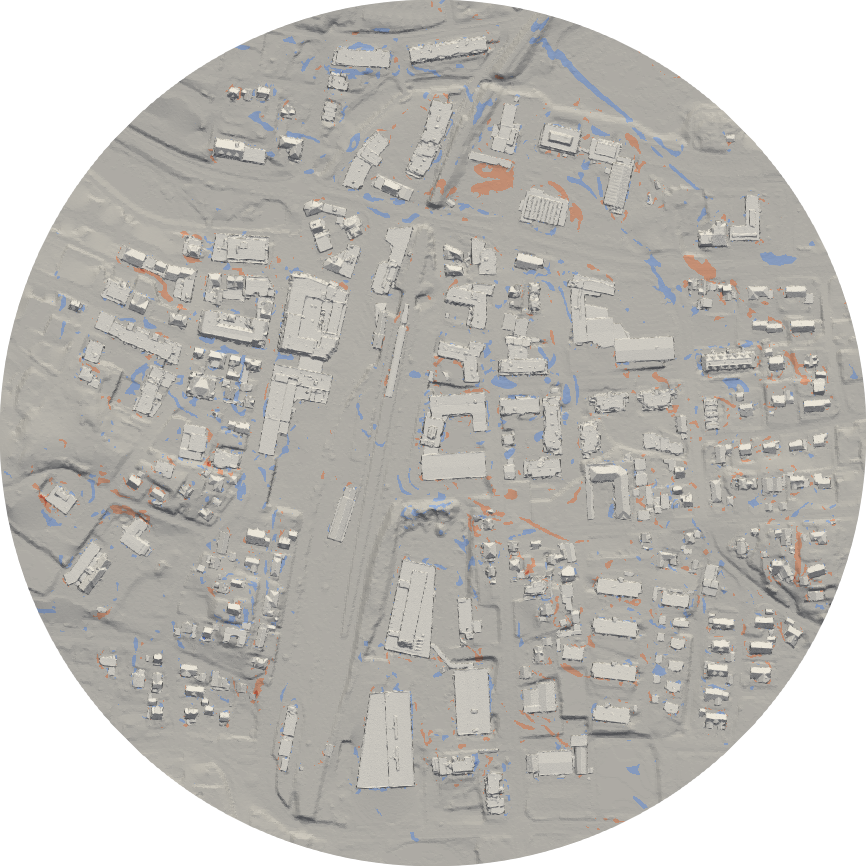}
        \caption{Drone photogrammetry model.}
    \end{subfigure}
    \begin{subfigure}{0.45\textwidth}
        \includegraphics[width=0.95\textwidth]{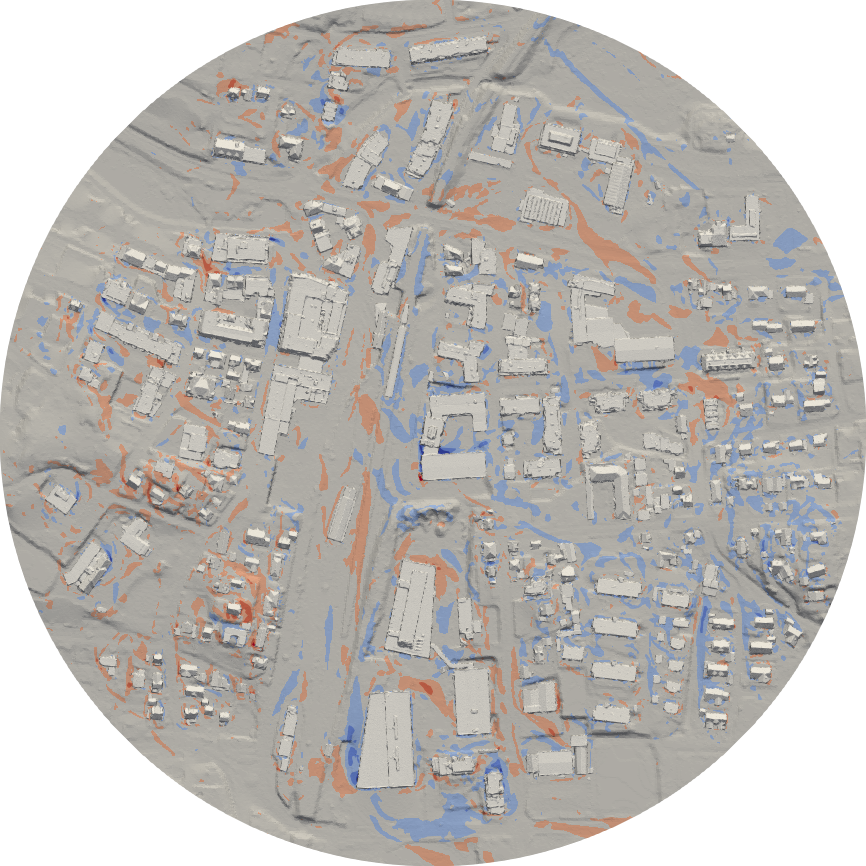}
        \caption{FKB database model.}
    \end{subfigure}
    
    \vspace{1pc}
    
    \begin{subfigure}{0.45\textwidth}
        \includegraphics[width=0.95\textwidth]{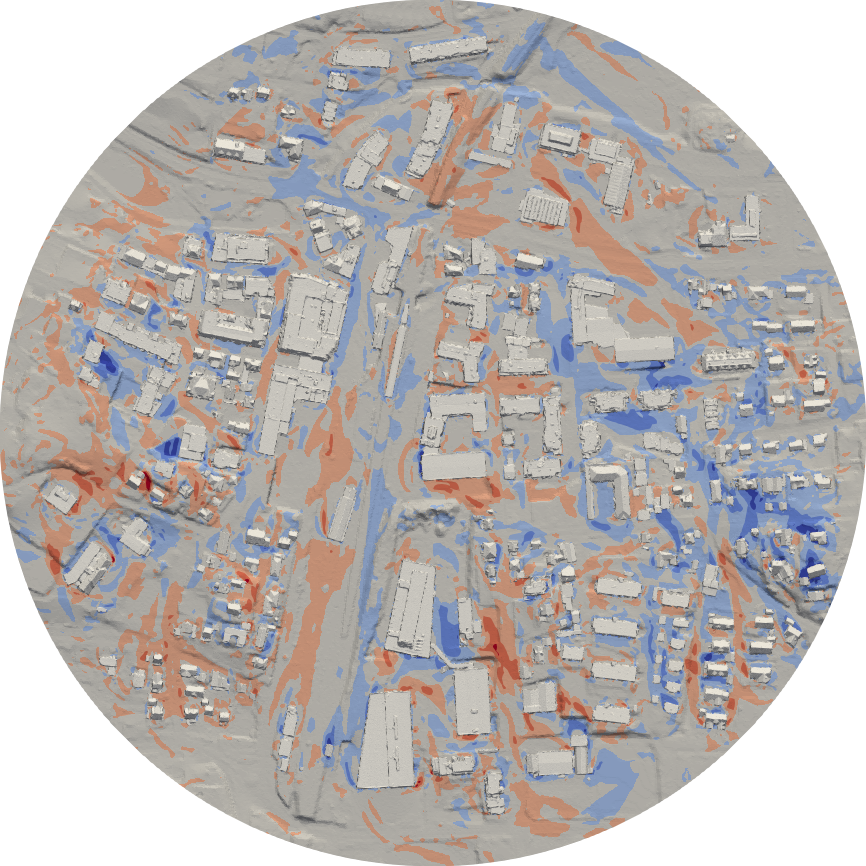}
        \caption{Footprint extrusion model.}
    \end{subfigure}
    
    \vspace{1pc}
    
    \includegraphics[width=0.4\textwidth]{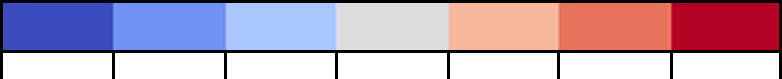}
    
    \begin{tikzpicture}
                \draw (-2.4, 0) node {$-3$};
                \draw (0, 0) node {$0$};
                \draw (2.4, 0) node {\makebox[0pt][l]{ $3$}\phantom{$-3$}};
                \draw (0, -0.5) node {$\Delta$C};
        \end{tikzpicture}
    \caption{Pedestrian wind comfort classification difference, $\Delta$C, to the reference model at \SI{2}{\meter} above ground.}
    \label{fig:cprojdiff}
\end{figure}

Figure \ref{fig:cproj} presents the pedestrian wind comfort maps at 2 m above ground using the four different building models, and Figure \ref{fig:cprojdiff} gives the classification difference to the reference case. 

A notable region by the high-rise building is classified as 'uncomfortable' in all the maps. The classification results from the wind frequently blowing from directions where the wind pull-down effect caused by the high-rise building is prominent. The wind is much weaker around the other lower buildings.

There are just a few small areas classified differently in the simulations using the Drone photogrammetry model compared to the reference case. A more noticeable area is classified differently in the FKB database model simulations. Most areas are classified with only one comfort class difference. The most considerable variation in the pedestrian wind comfort map to the reference case is using the Footprint extrusion model in the simulations. Here, some areas are even classified with three comfort classes difference. 

The most significant variations in the wind comfort maps are seen in the vicinity of the highest variations in the geometric building models themselves, see Figure \ref{fig:diff}.

Table \ref{tab:ssimC} presents the SSIM values of the pedestrian wind comfort maps (at \SI{2}{\meter} above ground level) compared to the reference case. The table confirms the variations seen in Figure \ref{fig:cproj} and Figure \ref{fig:cprojdiff} quantitatively, which is that the pedestrian wind comfort maps are most similar to the reference case when using the Drone photogrammetry model and the least when using the FKB database model. 

The SSIM values in Table \ref{tab:ssimC} are slightly lower than the values in Table \ref{tab:ssim}. It indicates that pedestrian wind comfort maps are more dependent on the geometric building model used than the pedestrian-level's wind velocity distribution. A possible explanation can be that the fields compared have been discretized into fewer classes in the pedestrian wind comfort maps (5 shades of gray) compared to the pedestrian-level's wind velocity distribution (12 shades of gray).

\begin{table}
\centering
\begin{tabular}{l l}
\hline
\textbf{Case compared to reference} & \textbf{SSIM}\\

\hline
Drone photogrammetry model & $ 0.92 $\\
FKB database model & $ 0.87 $\\
Footprint extrusion model & $ 0.80 $\\

\hline
\end{tabular}
\caption{Structural similarity of pedestrian wind comfort maps to reference map.}
\label{tab:ssimC}
\end{table}

\section{Conclusion}

This study has investigated the building geometry acquisition method's influence on the simulated wind field in an urban area focusing on pedestrian wind comfort. Four building model types of different acquisition methods and varying resemblance levels to the location were produced and tested in wind simulations. 

This study compares simulations without providing any validation measurements. Nevertheless, it is reasonable to conclude that the most detailed and accurate building models provide better foundations for more realistic and reliable wind simulations than using less precise building models. 

The building models that most closely represent the built environment's actual geometry are the Airborn LiDAR model and the Drone photogrammetry model, as they are based on highly precise remote sensing measurements without the oversimplification of the geometry. In contrast, the other building models have been simplified considerably, especially the FKB database model.  The building models' differences manifest themselves in the wind simulations as the simulations' variations are more significant near and downstream of the buildings with the most discrepancies to the reference model. 

Using the Airborne LiDAR model and the Drone photogrammetry model in the wind simulations yields quite similar results both in the wind velocity magnitudes at the pedestrian level and the pedestrian wind comfort maps. The wind simulations using the FKB database model showed minor differences to the simulations using the Airborn LiDAR model, while the results with the Footprint extrusion model showed significant differences. 

The simulations' variations are prominent enough to possibly affect building design decisions when planning the erection of new buildings in an urban environment. Using a FKB database model will likely result in the same conclusions as using a more detailed and accurate building model type. However, using a Footprint extrusion model instead may affect the decisions made. Therefore, care should be taken to ensure that the geometry model used in pedestrian wind comfort studies sufficiently resembles the actual built environment so that the wind simulations provide the best possible foundation for design decisions. 

Out of the four building model types compared in this study, the FKB database model is the best option for most practical applications of urban wind simulations. It provides similar results compared to using more detailed building models while having the possibility of being produced in a rather fast and automated manner. However, producing FKB database models naturally requires the building data to be stored in the database. This data sampling is usually achieved in a semi-automatic process requiring a degree of manual work. 

Further work includes evaluating more sites and possibly performing field measurements and or wind tunnel experiments for validation. A potential expansion is to study the influence of geometry acquisition methods on other applications of urban wind simulations, not just pedestrian wind comfort. Other applications include urban heat island and heat mitigation strategies, pollutant dispersion, and the prediction of extreme weather events. 

\section{Acknowledgements}

This research is part of the Future Energy Hub Project funded by The Norwegian Research Council (project no.: 280458), The University of Stavanger, and local industry partners. We want to thank Geodata, Geograf AS, and Statens Kartverk for providing the FKB database model, the drone photogrammetry point cloud, and the LiDAR point cloud, respectively. Data used to produce The Footprint extrusion model was acquired at OpenStreetMap, copyrighted and made available by OpenStreetMap contributors.






\bibliographystyle{apalike} 
\bibliography{sample.bib}






\end{document}